\documentclass[11pt,leqno,textwidth=10cm]{amsart}
\usepackage{amssymb,amsthm}
\usepackage{fourier}

\usepackage{mathrsfs}
\usepackage{mathtools}
\usepackage{abstract}
\usepackage{esint}
\usepackage{color}

\usepackage{etoolbox}
\makeatletter
\patchcmd{\@maketitle}{\newpage}{}{}{} 
\makeatother

\newtheoremstyle{fancy}{}{}{\itshape}{}{\textsc\bgroup}{.\egroup}{ }{}
\newtheoremstyle{fancy2}{}{}{\rm}{}{\textsc\bgroup}{.\egroup}{ }{}

\theoremstyle{fancy}
\newtheorem{Satz}{Theorem}[section]    
\newtheorem{cor}[Satz]{Corollary}
\newtheorem{lem}[Satz]{Lemma}
\newtheorem{prop}[Satz]{Proposition}
\newtheorem{thm}[Satz]{Theorem}

\newtheorem{named}{\name}

\newcommand{\name}{Proof of}

\theoremstyle{fancy2}

\newtheorem{rem}[Satz]{Remark}

\numberwithin{equation}{section}
\newcommand{\cref}[1]{Corollary~\ref{#1}}

\def\R{\mathbb R}

\renewcommand{\div}{\operatorname{div}}

\newcommand{\tr}{\operatorname{tr}}

\newcommand{\Chr}[3]{\Gamma^{#1}_{#2 #3}}
\newcommand{\hChr}[3]{\widehat{\Gamma}^{#1}_{#2 #3}}

\newlength\mtabskip\mtabskip=-1.25cm
  
 \def\mtabLong{long} 
 

\newcommand{\eq}[1]{\begin{equation}#1\end{equation}}

\newcommand{\alg}[1]{\begin{aligned}#1\end{aligned}}

\newcommand{\bc}{\begin{cases}}
\newcommand{\ec}{\end{cases}}

\newcommand{\p}[1]{\partial_{#1}}

\newcommand{\ab}[1]{|#1|}

\newcommand{\Abi}[1]{\|#1\|_{L^\infty}}

\newcommand{\Abk}[2]{\|#1\|_{H^{#2}}}

\newcommand{\abg}[1]{|#1|_g}

\newcommand{\Abbk}[2]{\VERT#1\VERT_{H^{#2}}}
\newcommand{\al}{\alpha}
\newcommand{\be}{\beta}
\newcommand{\ga}{\gamma}

\newcommand{\la}{\lambda}
\newcommand{\Si}{\Sigma}

\newcommand{\De}{\Delta}

\newcommand{\na}{\nabla}

\newcommand{\mcl}[1]{\mathcal{#1}}
\newcommand{\mcr}[1]{\mathscr{#1}}

\def\div{ \mbox{div}}

\def\R{\ensuremath {\mathbb R}}
\def\tr{\ensuremath {\mbox{tr}}}

\begin{document}


\title[Stable cosmological Kaluza-Klein Spacetimes ]{Stable cosmological Kaluza-Klein Spacetimes}
\author[V.~Branding, D.~Fajman,  K.~Kr\"oncke]{Volker Branding, David Fajman, Klaus Kr\"oncke}
\date{\today}


\address{Volker Branding\newline
Faculty of Mathematics, University of Vienna, Oskar Morgenstern Platz 1,1090 Vienna, Austria\\ \newline
Volker.Branding@univie.ac.at
}

\address{
David Fajman\newline
Faculty of Physics, University of Vienna, Boltzmanngasse 5,1090 Vienna, Austria\\ \newline
David.Fajman@univie.ac.at
}

\address{
Klaus Kr\"oncke\newline
Faculty of Mathematics, University of Hamburg, Bundesstrasse 55, Hamburg, Germany \newline
Klaus.Kroencke@uni-hamburg.de
}

\maketitle
\begin{abstract}
We consider the Einstein flow on a product manifold with one factor being a compact quotient of 3-dimensional hyperbolic space without boundary and the other factor being a flat torus of fixed arbitrary dimension. 
We consider initial data symmetric with respect to the toroidal directions.
We obtain effective Einsteinian field equations coupled to a wave map type and a Maxwell type equation by the Kaluza-Klein reduction. 
The Milne universe solves those field equations when the additional parts arising from the toroidal dimensions are chosen constant.
We prove future stability of the Milne universe within this class of spacetimes, which establishes stability of a large class of cosmological Kaluza-Klein vacua. {A crucial part of the proof is the implementation of a new gauge for Maxwell-type equations in the cosmological context, which we refer to as \emph{slice-adapted gauge}.}
\end{abstract}


\section{Introduction}

\subsection{Kaluza-Klein Spacetimes}
The classical approach to unification of general relativity with electromagnetism and more generally with gauge fields goes back to the original works of Kaluza and Klein. The Kaluza-Klein approach considers general relativity in 4+$n$ dimensions with spacetime factorizing as 
\eq{
M^{(4+q)}=M^{(4)}\times B,
}
where $M^{(4)}$ corresponds to the macroscopic spacetime and $B$ is a compact $q$-dimensional Riemannian manifold referred to as \emph{internal space}. The latter models compactified dimensions practically invisible to observers.\\
Identifying the ground state of Kaluza-Klein theory has been a long-standing open problem, which may be considered in different contexts. The terminology \emph{ground state} here refers to a stable fixed point of the Einstein flow. Original works show semiclassical instabilities in the case $M^{(4)}$ is equipped with the Minkowski metric \cite{Wi82}.  If results on the Einstein-Maxwell system \cite{BZ,Sp12}, which relate to the special case $B=S^1$ and $M^{(4)}$ being equipped with the Minkowski metric, are excluded, then mathematically rigorous nonlinear stability or instability of Kaluza-Klein spacetimes in the context of classical general relativity was unknown until recently. In a recent work Wyatt established stability of Kaluza-Klein spacetimes for the class of models, where $M^{(4)}$ carries the Minkowski metric and $B$ is a flat $q$-dimensional torus \cite{Wy17},
\eq{
\overline g_{KK} = \eta_{M^{(4)}}+g_{\mathrm{flat},\mathbb T^q}.
}
In the class of Kaluza-Klein spacetimes, the vacuum Einstein equations on $M^{(4+q)}$ reduce to an Einstein-wave map-Maxwell type system on $M^{(4)}$, which is shown to have the Minkowski metric as its stable ground state. From the perspective of classical general relativity this result justifies the interpretation of the corresponding higher-dimensional Kaluza-Klein background spacetime $(M^{(4+q)},\overline g_{KK})$ as the ground state of the generalized higher dimensional field equations.

\subsection{Cosmological spacetimes}
A prerequisite for the stability analysis of the class of Kaluza-Klein spacetimes with Minkowski space as their macroscopic part is the corresponding nonlinear stability result for the classical 4-dimensional vacuum Einstein equations \cite{CK92,LR}. In the class of asymptotically flat spacetimes Minkowski spacetime is the only solution known to be stable. The analogous problem for the Kerr family is still open. There is only one other spacetime in the class of solutions of the Einstein equations with vanishing cosmological constant that is known to be stable, which is the Milne model. This solution belongs to the class of cosmological spacetimes, i.e.~it has spatial slices with compact topology that carry a negative Einstein metric $\gamma$. The Milne model is future complete and past incomplete and its future nonlinear stability problem has been resolved in the vacuum setting by Andersson and Moncrief \cite{AnMo11}. This result covers also the higher-dimensional case, however, not in the sense of compactified dimensions. In analogy to the asymptotically flat case we ask for the natural ground state for Kaluza-Klein theory in the class of cosmological spacetimes. It will be shown in this paper that the generalized Kaluza-Klein spacetime arising from the Milne model, reading
\eq{
-dt^2+ \frac{t^2}9\gamma + g_{\mathrm{flat},\mathbb T^q},
}
is future nonlinearly stable for perturbations that are invariant under the isometry group of $g_{\mathrm{flat},\mathbb T^q}$. In the following, we call this invariance just $\mathbb T^q$-invariance.

\subsection{Main theorem}
\subsubsection{Result}
We first state a rough version of our main result. A detailed version will be given later in Theorem \ref{thm-1-version2}. The Sobolev norms used in the statement are defined with respect to the metric $\gamma$.
\begin{thm}
\label{thm-1}
Let $(M,\gamma)$ be a compact, negative, 3-dimensional Einstein manifold without boundary and Einstein constant $\mu=-\frac29$ and $g_{\mathrm{flat},\mathbb T^q}$ a flat metric on $\mathbb T^q$. Then there exists an $\epsilon>0$ such that for $\mathbb T^q$-invariant initial data $(g,k)$ on $M\times \mathbb T^q$ satisfying
\begin{align}
\left\|g-(\gamma+g_{\mathrm{flat},\mathbb T^q})\right\|_{H^4}+\left\|k-(1/9\cdot\gamma+0)\right\|_{H^3}<\epsilon
\end{align}
the corresponding maximal globally hyperbolic development under the Einstein vacuum equation is $\mathbb T^q$-invariant (hence a Kaluza-Klein spacetime), future-global in time and future complete. Moreover, there exists a foliation of the spacetime by almost CMC hypersurfaces such that the induced metric $g_t$, $t\in [1,\infty)$ converges in $H^4\times H^3$ after a natural rescaling to a product metric $c\cdot\gamma+g'_{\mathrm{flat},\mathbb T^q}$ where $c$ is a constant close to $1$ and $g'_{\mathrm{flat},\mathbb T^q}$ is a flat metric on $\mathbb T^q$ which is close to $g_{\mathrm{flat},\mathbb T^q}$.
\end{thm}
We formulate the detailed version of the main result in terms of rescaled variables, adapted to the evolution.
At first the symmetry assumption reduces Einstein's equation to a system of Einstein equations in $3+1$ dimension coupled to a Maxwell-type equation and to a wave map type equation. 
 To obtain the final reduced equations two rescalings are performed. The first is a conformal rescaling necessary to avoid regularity problems arising from the Kaluza-Klein reduction. The second rescaling uses the CMC-time function to obtain variables which are scale free and independent of the expansion. By using the CMC-time function, the conformal metric admits a foliation by CMC hypersurfaces. These hypersurfaces are not CMC anymore with respect to the physical metric but almost CMC which justifies the corresponding sentence in the main theorem.
 
   The detailed reduced system is given in \eqref{Co-Ha}-\eqref{ev-k}. We consider initial data sets consisting of a Riemannian metric $g$ on $M$, the trace-free part $\Si$ of the second fundamental form restricted to $M$, an $\R^q$-valued one-form $A$ (corresponding to the mixed terms of the metric on the product $M\times \mathbb T^q$) and its time derivative $\dot A$ as well as set of wave-type maps $\Phi$ (which is formally a map $\Phi:M\to \mathrm{GL}(q,\R)$ corresponding to a flat metric on $\mathbb T^q$) and their time derivatives $\dot \Phi$ fulfilling the reduced constraint equations \eqref{Co-Ha} -- \eqref{Co-Mo}.

 \begin{rem}The one-form $A$ is coupled to the full system via a Maxwell-type equation (see \eqref{maxwell-eq} below). 
To obtain a suitable solution theory for this equation, we need to impose a gauge condition, e.g.\ the Lorentz gauge, which turns \eqref{maxwell-eq} into a hyperbolic equation. In the main theorem, we have imposed the Lorentz gauge and the initial data $(A,\dot A)$ is meant as initial data with respect to this hyperbolic equation. However, to control the long-time behaviour of \eqref{maxwell-eq}, a different gauge turned out to be more suitable, see Subsection \ref{section-sl-ad-gauge} below.
\end{rem}

\begin{rem}
The Kaluza-Klein reduced Einstein equations restrict all possible perturbations of the background to those which preserve the isometry group of the flat torus. 
This, however, still allows that at each point in the macroscopic space the torus 
(which is the internal space at this point) may evolve within the class of flat tori. 
\end{rem}

\begin{rem}
By the conformal rescaling we perform in Section \ref{conf_resc} of this paper, the Riemannian metrics $\bar{g}$ of the spatial hypersurfaces satisfy $\tau^2\bar{g}\to \det(\Phi_{\infty})^{-\frac{1}{2}}\gamma$ as $\tau\to 0$. Here, $\tau$ represents 
the mean curvature of the hypersurfaces. 
Moreover, as $\Phi_{\infty}$ is constant, this limit metric is also negative Einstein but with a possible different Einstein constant. It is interesting to note that the macroscopic geometry encoded in $\bar{g}$ is affected by the geometry of the internal space through the above rescaling.
\end{rem}

\begin{rem}
Our main result can also be applied to the stability analysis of classical vacua in string theory since toroidal compactifications are often employed as toy models here.
\cite[Chapter 8]{Po05}.
\end{rem}

\begin{rem}
In order to connect the present result to the existing literature we would like to point that another stability result for the Einstein flow (with positive cosmological constant) holds for the \emph{deSitter solution} and corresponding counterparts with other spatial topologies \cite{Ri08}. The same analysis could also be performed in the context of compactified dimensions. However, the main difference between the Milne model considered here and deSitter space lies in the fact that the presence of a positive cosmological constant causes an accelerated expansion while the Milne model and perturbations of it only experience linear expansion. In consequence, as shown in \cite{Ri08}, the analysis of the stability of deSitter space localizes in space and the topology of the spatial slices becomes irrelevant in the analysis (as long as a suitable background solutions exists). This effect is not present in the case of the Milne universe, which makes the particular approach by Andersson and Moncrief \cite{AnMo11} necessary. For more details in this regard we refer to \cite{Kr16} for a presentation of the respective conformal structures.
\end{rem}

We comment in the following on some technical aspects of the stability proof.
\subsubsection{Gauging Kaluza-Klein fields: The slice-adapted gauge}\label{section-sl-ad-gauge}
A standard gauge for a vector potential that is used to consider Maxwell-type equations is the Lorentz gauge $\nabla^{\mu}A_{\mu}=0$. One obtains a nonlinear wave equation of second order on $A$. However, it turns out to be surprisingly difficult to analyze this equation in the present context and to construct a natural energy which yields optimal bounds for the decay of the perturbation. A source of this difficulty may arise from the fact that the vector potential $A$ is not determined by the Lorentz gauge as this gauge is preserved by transformations $A\mapsto A+df$ if $\Box f=0$ and thus has infinitely many degrees of freedom. To overcome this problem, we choose a gauge which is \emph{adapted} to a foliation of the spacetime by spacelike hypersurfaces and which uniquely determines $A$:
We demand that the spatial components of $A$, $\omega$, associated to this foliation are divergence-free and orthogonal to the kernel of the Hodge Laplacian and that the time component of $A$, $\Psi$, 
regarded as a function of the spacetime has vanishing integral on each spatial slice
\begin{align}
\mathrm{div}_g\omega=0 ,\qquad
\qquad \omega\perp\ker(\Delta_{H}), \qquad\int_M \Psi dV_g=0.
\end{align}

In this gauge, the Maxwell equation is a wave equation on the spatial part of $A$ coupled to an elliptic equation for its time component.
Details are provided in Lemma \ref{gauge} and Proposition \ref{form-eq-1}.  To the best of our knowledge, such a gauge has not been used in the context of related problems so far. However, the slice-adapted gauge can be applied to Maxwell-type equations on other spacetimes with compact spatial hypersurfaces (e.g.\ on the deSitter space).

Recently, several gauges for the analysis of the Einstein equations were introduced, which are of elliptic or parabolic nature \cite{RoSp18,RoSp18-b}. We would like to point out that the latter do not have any relation to the slice-adapted gauge which we introduce in this paper.    
\subsubsection{Regularity aspects and the momentum constraint}
Another interesting aspect of the Kaluza-Klein reduced system is the fact that the momentum constraint, which is not explicitly used in controlling the perturbation in the pure 3+1-dimensional vacuum stability proof, does play an important role in the present problem in the following sense. 
Below, we will use energies that control the $H^4$-norm of an evolving metric $g$ (in terms of a fixed background metric) and the $H^{3}$-norm of the tracefree part $\Sigma$ of the second fundamental form. However, when differentiating the energies for the perturbation of the fields generated by the internal space, one obtains $4$ derivatives of $\Sigma$ and $3$ derivatives of its time derivative. Those in turn can not be controlled by the $H^{3}$-norm of $\Sigma$ and the $H^{2}$-norm of its time derivative, respectively. A closer analysis however reveals that these terms only appear as third derivatives of $\div_g\Sigma$ and second derivatives of $\partial_T\div_g\Sigma$. Replacing those terms using the momentum constraint improves the regularity by one order and closes the estimate.
\subsection{Related systems}
Theorem 1 has some immediate consequences for related systems and in particular automatically implies the following results.
\subsubsection{Einstein-Maxwell-Dilaton system}
In the special case of $B=\mathbb T^q=S^1$ the 5-dimensional $U(1)$-symmetric vacuum field equations with $S^1$ being the symmetry direction reduce to the 4-dimensional Einstein-Maxwell-Dilaton system \cite{Ow97}. This implies in particular the following corollary.

\begin{cor}
The Milne model is future stable as a trivial solution to the Einstein-Maxwell-Dilaton system. 
\end{cor}
We use the terminology \emph{trivial solution} in the sense that it is actually a solution to the Einstein vacuum equations. \\

In the case that the field $\Phi$ is given by the identity map its equation
of motion is trivially satisfied and does not contain any geometric information.
In this setup we obtain a new system that is formally equivalent to the classical Einstein-Maxwell system. This implies

\begin{cor}
The Milne model is future stable as a trivial solution to the Einstein-Maxwell system. 
\end{cor}

\subsubsection{Brans-Dicke theory}
Another well-known system that is captured by our main result is the Brans-Dicke model of general relativity.
This system is obtained by setting the one-forms \(A\) to zero.
The Brans-Dicke model couples pure gravitation with a scalar field in which the 
value of the scalar field can be interpreted as a dynamical version of Newton's gravitational constant,
see \cite{Ow97} for more details.

\begin{cor}
The Milne model is future stable as a trivial solution to the Brans-Dicke model. 
\end{cor}

\subsubsection{U(1)-symmetric spacetimes}
There is a third relation of Theorem \ref{thm-1} with previously considered models, where in this specific case the present result can be considered as a higher-dimensional analog. In their work on the stability of certain Bianchi type-III models Choquet-Bruhat and Moncrief consider spatial topologies of the form $\Sigma\times S^1$, where $\Sigma$ is a closed two-dimensional higher genus surface \cite{CM01}. The background solution being investigated is $-4dt^2+2 t^2\sigma_\Si+dx^2$, where $\sigma_\Si$ is a metric of constant negative scalar curvature on $\Si$. They prove future stability of this solution considered within the set of solutions to the 
4-dimensional vacuum Einstein equations obeying a $U(1)$-symmetry in the $S^1$ direction. By a Kaluza-Klein reduction this symmetric system is equivalent to the 2+1-dimensional Einstein equations on $\mathbb R\times \Sigma$ with a source term given by a massless scalar field. In a way this can be seen as an analogue to the problem considered in the present work, where the Kaluza-Klein fields are replaced by a single massless scalar field. However, the approach of Choquet-Bruhat and Moncrief does not carry over to higher dimensions as it relies on the particular features of the 2+1-dimensional geometric setting. Those are for instance the existence of a monotone $L^2$-energy and the usability of the momentum constraint to control the trace-free part of the second fundamental form. In 3+1-dimensions these methods are not available and need to be replaced by the energies provided by Andersson-Moncrief \cite{AnMo11}. Nevertheless, the structure of a torus bundle over a negatively curved compact Riemannian manifold is present in both cases. The result in this paper implies that those geometries are stable under the Einstein flow irrespective of the low dimensional features used in \cite{CM01}.
\subsubsection{Higher-dimensional backgrounds}
Finally, we mention that by the methods used in this paper, one can also prove nonlinear stability of a higher-dimensional Kaluza-Klein Milne model
\begin{align}
-dt^2+\frac{t^2}{m^2}\gamma+g_{\mathrm{flat},\mathbb{T}^q}
\end{align}
under the same class of perturbations.
Here, $\gamma$ is a negative Einstein metric with Einstein constant $-(m-1)/m^2$ on
a compact $m$-dimensional manifold. 
In higher dimensions, the conformal behaviour of the Maxwell-type equation yields a faster decay of $F_{\mu\nu}$ and improves the energy estimates.

\subsection{Organization of the paper}
This paper is organized as follows. In Section \ref{sec : prel} notations are introduced as well as the rescaling of the macroscopic geometry and several auxiliary quantities.
In Section \ref{sec : KKR} we perform the Kaluza-Klein reduction and derive the reduced Einstein-wave map-Maxwell system.
In Section \ref{sec : EMT} we compute the energy-momentum tensor in the reduced Einstein equations in terms of the fields generated by the internal space and introduce norms to estimate them. 
Section \ref{sec : hyp} derives energy estimates for all evolution equations individually and thereby constitutes the core step of the stability analysis. Section \ref{sec : ell} presents the elliptic estimates for the macroscopic lapse function and the shift vector field. 
Section \ref{sec : pro} presents the proof of the main theorem and Section \ref{sec : rel} presents all related systems listed above for which our stability analysis of the Milne model applies.

\subsection*{Acknowledgements}
We thank the anonymous referee for his remarks and suggestions that helped to improve the paper. D.F.~has been supported by the Austrian Science Fund (FWF) project P29900-N27 \emph{Geometric Transport equations and the non-vacuum Einstein flow}. V.B.~ gratefully acknowledges the support of the Austrian Science Fund (FWF) 
through the project P30749-N35 \emph{Geometric variational problems from string theory}.

\section{Preliminaries} \label{sec : prel}
\subsection{Notation} 
Throughout this paper, $M$ is a compact manifold eventually equipped with different Riemannian metrics and $I\subset \R$ is an open interval.
In this paper, the appearing Lorentzian metrics on $\widetilde{M}=I\times M$ will be denoted by $h,$ and the associated covariant derivative will be denoted by $\nabla$. The wave operator associated to $h$ is defined with the sign convention such that $\Box=\tr_{h}\nabla^2$.
In this paper, we will sometimes also denote Lorentzian metrics by $\tilde{h},\bar{h},\hat{h}$ and the associated covariant derivatives and wave operators will be denoted by $\widetilde{\nabla},\overline{\nabla},\hat{\nabla}$ and $\widetilde{\Box},\overline{\Box}$ and $\hat{\Box}$, respectively. Riemannian metrics on $M$ will be denoted by $g,\tilde{g}$ and the associated covariant derivatives will be denoted by $D,\widetilde{D}$, respectively. The Laplacian of $g$ is defined as $\Delta=\tr_gD^2$ and the volume form will be denoted by $dV_g$. The exterior derivative acting on differential forms on $M$ is denoted by $d$ and the formal adjoint with respect to $g$ is $d^*$. The Hodge-Laplacian acting on differential forms is then $\Delta_H=d^*d+dd^*$.
The Lie-derivative of a tensor $T$ in the direction of a vector field $X$ will be denoted by $\mathcal{L}_XT$. Throughout this paper, Greek indices $\alpha,\beta,\gamma,\ldots$ will denote spacetime coordinates on $I\times M$ and Latin indices $i,j,k,\ldots$ will denote coordinates on $M$. The coordinates on the torus $\mathbb{T}^q$ will be denoted by $m,n,p,\ldots$. The index $0$ will either refer to a time coordinate or to a timelike vector field. Its meaning will be clarified in the subsection where it is used.

\subsection{The macroscopic spatial background geometry}
In what follows we consider $M$ equipped with a negative Riemannian Einstein metric $\gamma$ with $\mathrm{Ric}[\ga]=-\frac{2}{9}\ga$ fixed once and for all. 
The Einstein operator $\Delta_E$ associated with $\ga$ acting on symmetric $2$-tensors, $\Delta_E\equiv-\Delta -2\mathring{R}$, has trivial kernel, i.e.~$\ker \Delta_E=\{0\}$.
This fact is relevant for the features of the natural energy associated with $\Delta_E$. This has been discussed in \cite{AnFa17} and is mentioned here for the sake of completeness.

\subsection{Geometric formalism for the evolving spacetime}\label{formalism}
In the following sections, we will study the evolution of a 3+1-dimensional Lorentzian metric $\tilde{h}$ (more precisely of its rescaled version $h$ introduced below). For this purpose, we will now introduce some geometric quantities that will be used throughout the paper. 
In the ADM formalism, $\tilde{h}$ is written as
\begin{align}\label{ADM-ansatz}
\tilde{h}=-\widetilde{N}^2d\tau^2+\tilde{g}_{ij}(dx^i+\widetilde{X}^id\tau)\otimes(dx^j+\widetilde{X}^jd\tau),\qquad \tau\in (-\infty,0)
\end{align}
and the tracefree part of the second fundamental form of the hypersurfaces $\left\{\tau=const\right\}$ is denoted by $\widetilde{\Sigma}$. Here we assume that these hypersurfaces all have constant mean curvature and that the mean curvature of $\left\{\tau=const\right\}$ is $\tau$.
We define rescaled quantities $g,N,\Sigma,X$ by
\begin{align}\label{rescaling1}
g_{ij}=\tau^2\tilde{g}_{ij},\qquad N=\tau^2\widetilde{N},\qquad \Sigma_{ij}=\tau\widetilde{\Sigma}_{ij},\qquad X^i=\tau\widetilde{X}^i
\end{align}
and a rescaled time $T$ via
\begin{align}\label{rescaling2}
\tau=\tau_0\cdot e^{-T},\qquad T\in (-\infty,\infty),\quad\tau_0<0 \text{ is fixed.} 
\end{align}
It is easily seen that with respect to this new time coordinate, the above Lorentzian metric is given by
\begin{align}\label{metric-h}
\tilde{h}=(\tau_0)^{-2}e^{2T}(-N^2dT^2+g_{ij}(dx^i-X^idT)\otimes(dx^j-X^{j}dT))=:(\tau_0)^{-2}e^{2T}\cdot h.
\end{align}

Let $\Pi$ be the second fundamental form of the slice $\left\{T\equiv const\right\}$ with respect to the Lorentzian metric $h$. Then one can show that
\begin{align}
\Pi=-\Sigma+N^{-1}(1-N/{3})g.
\end{align}
The future-directed timelike unit normal of the hypersurfaces  $\left\{T\equiv const\right\}$ with respect to $h$  is
\begin{align}\label{e_0}
e_0=N^{-1}(\partial_T+X).
\end{align}
We use $e_0$ to split $1$-forms on $\widetilde{M}$ described in the following. For $A\in \Omega^1(\widetilde{M})$, we define a function $\Psi\in  C^{\infty}(\widetilde{M})$ and a time-dependent family of one-forms $\omega\in C^{\infty}(I,\Omega^1(M))$ by $\Psi:=A(e_0)$ and $\omega(\partial_i)=A(\partial_i)$. 
Throughout the paper, we will view any $\omega\in  C^{\infty}(I,\Omega^1(M))$ as an element in $\Omega^1(\widetilde{M})$ by demanding $\omega(e_0)=0$. This allows us to write the above splitting as $A=\omega+\Psi e_0^*$ where $e_0^*\in \Omega^1(\widetilde{M})$ is the dual of $e_0$. We compute the connection coefficients for the rescaled Lorentzian metric $h$. Using the Koszul formula, one shows
\begin{equation}\label{christoffel}
\Gamma(h)^{0}_{00}=\Gamma(h)_{i0}^0=0,\quad \Gamma(h)_{00}^i=g^{ij}N^{-1}{\partial_jN},\quad \Gamma(h)_{ij}^0=-\Pi_{ij},\quad \Gamma(h)_{i0}^j=-g^{jl}\Pi_{li},\quad \Gamma(h)_{ij}^k=\Gamma(g)_{ij}^k,
\end{equation}
where $i,j,k$ are coordinates on $M$ and the index $0$ refers to the vector field $e_0$ given in \eqref{e_0}.
The following lemma is technically relevant for computations performed further below.
\begin{lem}\label{variations}
We have
\begin{equation}\begin{split}
\mathcal{L}_{e_0}g&
=-2\Pi,\\
[\mathcal{L}_{e_0},\mathrm{div}_g]\eta&=2\langle\Pi,D\eta\rangle+\langle D\log N,\mathcal{L}_{e_0}\eta\rangle+\langle S,\eta\rangle +2\langle\Pi,D\log N\otimes\eta\rangle-\mathrm{tr}_g\Pi\langle D\log N,\eta\rangle,\\
[\mathcal{L}_{e_0},\Delta_g]f&=2\langle\Pi,D^2f\rangle+\langle D\log N,D\partial_{e_0}f\rangle+\langle S,Df\rangle +2\langle\Pi,D\log N\otimes Df\rangle-\mathrm{tr}_g\Pi\langle D\log N,Df\rangle	
\end{split}
\end{equation}
for all $f\in C^{\infty}(\widetilde{M})$ and $\eta\in C^{\infty}(I,\Omega^1(M))$.
Here, $S=2\mathrm{div}_g\Pi-D\tr_g\Pi$.
\end{lem}

\begin{proof}
At first, we compute
\begin{align}
\mathcal{L}_{e_0}g
=\mathcal{L}_{N^{-1}(\partial_T+X)}g=N^{-1}\mathcal{L}_{(\partial_T+X)}g
=-2\Pi.
\end{align}	
Let $\left\{\partial_1,\partial_2,\partial_3\right\} $ be local coordinate fields on $M$ such that $D_{\partial_i}\partial_j=0$ at some fixed point $p$ and with respect to a fixed metric $g_{t_0}$. We then extend these local vector fields to elements in $ C^{\infty}(I,\mathfrak{X}(M))$ by defining $\partial_i(t)=\varphi_t^*\partial_i(t_0)$, where $\varphi_t\in \mathrm{Diff}(M)$ is generated by $-X$. Then by construction, $[\partial_T+X,\partial_i]=0$ and therefore, $[e_0,\partial_i]=\frac{\partial_iN}{N}e_0$. 	
Then at the point $(t,p)$, we compute
\begin{equation}
\begin{split}
\partial_{e_0}\Gamma_{ij}^k
&=\frac{1}{2}g^{kl}\partial_{e_0}(\partial_ig_{jl}+\partial_jg_{il}-\partial_lg_{ij})\\
&=\frac{1}{2}g^{kl}(\partial_i\partial_{e_0}g_{jl}+\partial_j\partial_{e_0}g_{il}-\partial_l\partial_{e_0}g_{ij})
+\frac{1}{2}g^{kl}([e_0,\partial_i]g_{jl}+[e_0,\partial_j]g_{il}-[e_0,\partial_l]g_{ij})
\\
&=\frac{1}{2}g^{kl}(\partial_i(\mathcal{L}_{e_0}g)_{jl}+\partial_j(\mathcal{L}_{e_0}g)_{il}-\partial_l(\mathcal{L}_{e_0}g)_{ij})\\ &\quad
+\frac{1}{2}g^{kl}(\frac{\partial_iN}{N}\partial_{e_0}g_{jl}+\frac{\partial_jN}{N}\partial_{e_0}g_{il}-\frac{\partial_lN}{N}\partial_{e_0}g_{ij})
\\
&=\frac{1}{2}g^{kl}(D_i(\mathcal{L}_{e_0}g)_{jl}+D_j(\mathcal{L}_{e_0}g)_{il}-D_l(\mathcal{L}_{e_0}g)_{ij})
\\ &\quad
+\frac{1}{2}g^{kl}(\frac{\partial_iN}{N}\mathcal{L}_{e_0}g_{jl}+\frac{\partial_jN}{N}\mathcal{L}_{e_0}g_{il}-\frac{\partial_lN}{N}\mathcal{L}_{e_0}g_{ij}).
\\
\end{split}
\end{equation}
Therefore, we obtain
\begin{equation}
\begin{split}
\mathcal{L}_{e_0}(\mathrm{div}_g\eta)&-\mathrm{div}_g(\mathcal{L}_{e_0}\eta)
=(\partial_{e_0}g^{ij})(\partial_{i}\eta_j-\Gamma_{ij}^k\eta_k)+g^{ij}[e_0,\partial_i]\omega_j-
g^{ij}(\partial_{e_0}\Gamma_{ij}^k)\eta_k\\
&=2g^{ki}g^{lj}\Pi_{kl}(\partial_{i}\eta_j-\Gamma_{ij}^k\eta_k)
+g^{ij}\frac{\partial_iN}{N}\partial_{e_0}\eta_j
+ g^{ij}g^{kl}(D_i\Pi_{jl}+D_j\Pi_{il}-D_l\Pi_{ij})\eta_k\\&\quad
+ g^{ij}g^{kl}(\frac{\partial_iN}{N}\Pi_{jl}+\frac{\partial_jN}{N}\Pi_{il}-\frac{\partial_lN}{N}\Pi_{ij})\eta_k.
\end{split}
\end{equation}
The third formula follows from the second and the fact that $[\mathcal{L}_{e_0},D]f=0$.
\qedhere 
\end{proof}

\section{Kaluza-Klein reduction}\label{sec : KKR}
In this section we perform the Kaluza-Klein reduction beginning with the physical Lorentzian metric on the full spacetime.
\subsection{Kaluza-Klein metrics}
Following \cite[p.~653]{CH09}  we consider the Kaluza-Klein ansatz for the Lorentzian metric $\hat h$ on $\mathbb R\times M\times \mathbb T^q$, where $q\geq 1$ denotes the dimension of internal space,
\eq{
\hat h_{AB}\hat{\theta}^A\hat{\theta}^B=\bar{h}_{\alpha\beta}\theta^\alpha\theta^\beta+\Phi_{mn}(\theta^m+A_{\alpha}^m\theta^\alpha)(\theta^n+A_{\beta}^n\theta^\beta).
}
Here, $\bar{h}_{\alpha\beta}$ denotes a Lorentzian metric on $\mathbb R\times M$, $\{\Phi_{mn}\}$ is a set of functions on $M$ and $A_\al^i$  is an $\mathbb{R}^q$-valued 1-form on $M$. Moreover, $\theta^\al$ and $\theta^m$ are suitable co-frames on $M$ and $\mathbb T^q$, respectively. We obtain for the macroscopic part of the Ricci tensor (cf.~\cite[p.~ 659 eq.~(5.2)]{CH09}) 
\eq{
\hat R_{\alpha\beta}=\overline{R}_{\alpha\beta}-\frac12 F_{m,\beta}^\mu F_{\alpha\mu}^m-\frac14\left(\Phi^{mq}(\overline{\nabla}^2_{\alpha\beta}+\overline{\nabla}^2_{\beta\alpha})\Phi_{mq}+\overline{\nabla}_\alpha \Phi^{mq}\overline{\nabla}_\beta \Phi_{mq}\right).
}
Here,  $\left\{\Phi^{mn}\right\}_{1\leq m,n\leq q}$ is the inverse of the matrix $\left\{\Phi_{mn}\right\}_{1\leq m,n\leq q}$ and $F$ is the curvature of $A$ given by $F=dA$ or equivalently $F_{\mu\nu}=\p \mu A_\nu-\p\nu A_\mu$.
Imposing Einstein equations on $\hat R_{\alpha\beta}$ yields
\eq{\alg{\label{einstein-eq}
\overline{R}_{\alpha\be}-\frac12 \overline{R} \bar{h}_{\al\be}&=\frac12 F_{m,\beta}^\mu F_{\alpha\mu}^m+\frac14\left(\Phi^{mq}(\overline{\nabla}^2_{\alpha\beta}+\overline{\nabla}^2_{\beta\alpha})\Phi_{mq}+\overline{\nabla}_\alpha \Phi^{mq}\overline{\nabla}_\beta \Phi_{mq}\right)\\
&\quad- \frac12\left(\frac12 F_{m,\beta}^\mu F_{\mu}^{m\be}+\frac12\left(\Phi^{mq}\overline{\Box}\Phi_{mq}+\frac12\overline{\nabla}_\alpha \Phi^{mq}\overline{\nabla}_\beta \Phi_{mq}\right)\right)\bar{h}_{\alpha\beta}=: \overline{T}_{\alpha\beta}[\Phi,F].
}}
In particular, $\overline{T}[\Phi,F]$ determines the matter source terms in the effective 3+1-dimensional Einstein equations. The remaining parts of the Einstein vacuum equation yield the equations of motion for the fields $F$ and $\Phi$, cf.~\cite[p.~ 659, eq.~(5.3, 5.4)]{CH09}. Those equations read
\begin{align}
\frac12 \overline{\nabla}_\lambda F_{l,\alpha}^{\lambda}+\frac14 F_{l,\alpha}^{\lambda}\Phi^{mp}\overline{\nabla}_{\lambda}\Phi_{np}=0,\\
-2\overline{\Box}\Phi_{mn}-\Phi^{pq}\overline{\nabla}_{\alpha}\Phi_{pq}\overline{\nabla}^{\alpha}\Phi_{mn}+2\Phi^{pq}\overline{\nabla}_{\alpha}\Phi_{mp}\overline{\nabla}^{\alpha}\Phi_{nq}+F_{m,\al\be}F_n^{\al\be}=0.
\end{align}
\subsection{Conformal rescaling}\label{conf_resc}
From an analytical point of view the second order terms of $\Phi$ on the right-hand side of \eqref{einstein-eq} are problematic. 
In this section, we therefore perform a standard conformal rescaling of the metric that yields an equivalent system that has a better analytic structure.
Let us recall some standard transformation formulas.
If  $\bar{h}=e^{2u}\tilde{h}$, we have
\begin{equation}
\begin{split}
\bar{R}_{\alpha\beta}&=\widetilde{R}_{\alpha\beta}-(n-2)(\widetilde{\nabla}_{\alpha\beta}^2u-\widetilde{\nabla}_{\alpha}u\widetilde{\nabla}_{\beta}u)-(\widetilde{\Box}u+(n-2)|\widetilde{\nabla}u|^2)\tilde{h}_{\alpha\beta},\\
\bar{R}&=e^{-2u}(\tilde{R}-2(n-1)\widetilde{\Box}u-(n-2)(n-1)|\widetilde{\nabla}u|^2),\\
\overline{\nabla}^2_{\alpha\beta}f&=\widetilde{\nabla}_{\alpha\beta}^2f-\widetilde{\nabla}_{\alpha}u\widetilde{\nabla}_{\beta}f-\widetilde{\nabla}_{\beta}u\widetilde{\nabla}_{\alpha}f+\tilde{h}^{\lambda\mu}\widetilde{\nabla}_{\lambda}u\widetilde{\nabla}_{\mu}f\tilde{h}_{\alpha\beta},\\
\overline{\Box}f&=e^{-2u}(\widetilde{\Box} f+(n-2)\tilde{h}^{\lambda\mu}\widetilde{\nabla}_{\lambda}u\widetilde{\nabla}_{\mu}f),
\end{split}
\end{equation}
where $n$ is the dimension of the spacetime.
As a conformal factor, we set
\begin{align}
u=c\cdot \log(\det \Phi)
\end{align}
with a constant $c$ whose value is to be determined. We have
\begin{align}
\widetilde{\nabla}_{\alpha}u=c\cdot\Phi^{mn}\widetilde{\nabla}_{\alpha}\Phi_{mn},
\qquad \widetilde{\nabla}^2_{\alpha\beta}u=c(\Phi^{mn}\widetilde{\nabla}^2_{\alpha\beta}\Phi_{mn}+\widetilde{\nabla}_{\alpha}\Phi^{mn}\widetilde{\nabla}_{\beta}\Phi_{mn}),
\end{align}
where we used that for a matrix \(A\) we have
\(\partial_\alpha\det A=\det A~\tr(A^{-1}\partial_\alpha A)\).
We now put $c=-1/4$, $n=4$ and the relation between the physical metric $\bar{h}$ and the conformal metric $\tilde{h}$ is $\bar{h}=e^{2u}\tilde{h}=\frac{1}{\sqrt{\det\Phi}}\tilde{h}$. Then for $\tilde{h}$, $F$ and $\Phi$, the equations read
\begin{align}\label{maxwell-eq}
 \widetilde{\nabla}_{\lambda} F_{l,\alpha}^{\lambda}&=-\frac{1}{2} F_{l,\alpha}^{\lambda}\Phi^{mp}\widetilde{\nabla}_{\lambda}\Phi_{np},\\
 \label{wave-eq}
\widetilde{\Box}\Phi_{mn}&=\tilde{h}^{\mu\lambda}\Phi^{pq}\widetilde{\nabla}_{\mu}\Phi_{mp}\widetilde{\nabla}_{\lambda}\Phi_{nq}+\frac{1}{2}\sqrt{\det\Phi}F_{m,\mu\lambda}F_{n,\gamma\delta}
\tilde{h}^{\mu\gamma}\tilde{h}^{\lambda\delta}
\end{align}
and 
\begin{equation}
\begin{split}
\label{conformal system}
\widetilde{R}_{\alpha\beta}-\frac{1}{2}\widetilde{R}\tilde{h}_{\alpha\beta}&=
\frac{1}{2}\sqrt{\det \Phi}[F_{m,\mu\beta}F_{\alpha\lambda}^m\tilde{h}^{\mu\lambda}-\frac{1}{2}F_{m,\mu\lambda}F^m_{\rho\nu}\tilde{h}^{\mu\nu}\tilde{h}^{\lambda\rho}\tilde{h}_{\alpha\beta}+\frac{1}{4}\Phi^{mn}F_{m,\mu\lambda}F_{n,\rho\nu}\tilde{h}^{\mu\rho}\tilde{h}^{\lambda\nu}\tilde{h}_{\alpha\beta}]\\
&\quad -\frac{1}{4}\widetilde{\nabla}_{\alpha}\Phi^{mn}\widetilde{\nabla}_{\beta}\Phi_{mn}
+\frac{1}{8}\Phi^{pq}\widetilde{\nabla}_{\alpha}\Phi_{pq}\Phi^{mn}\widetilde{\nabla}_{\beta}\Phi_{mn}
+\frac{1}{4}
\Phi^{pq}\widetilde{\nabla}_{\mu}\Phi_{mp}\Phi^{mn}\widetilde{\nabla}_{\lambda}\Phi_{nq}\tilde{h}^{\mu\lambda}\tilde{h}_{\alpha\beta}\\
&\quad+\frac{3}{8}\tilde h^{\mu\lambda}\widetilde{\nabla}_{\mu}\Phi^{mn}\widetilde{\nabla}_{\lambda}\Phi_{mn}\tilde{h}_{\alpha\beta}-\frac{1}{16}\tilde{h}^{\mu\lambda}
\Phi^{pq}\widetilde{\nabla}_{\mu}\Phi_{pq}\Phi^{mn}\widetilde{\nabla}_{\lambda}\Phi_{mn}\tilde{h}_{\alpha\beta}.
\end{split}
\end{equation}

\subsection{Macroscopic Einstein equations}
\eqref{conformal system} are the \emph{effective macroscopic Einstein equations}. Their energy-momentum tensor, arising from the geometry of the internal space, takes the form
\eq{\alg{
\tilde T_{\al\be}[A,\Phi]&=
\frac{1}{2}\sqrt{\det \Phi}[F_{m,\mu\beta}F_{\alpha\lambda}^m\tilde{h}^{\mu\lambda}-\frac{1}{2}F_{m,\mu\lambda}F^m_{\rho\nu}\tilde{h}^{\mu\nu}\tilde{h}^{\lambda\rho}\tilde{h}_{\alpha\beta}+\frac{1}{4}\Phi^{mn}F_{m,\mu\lambda}F_{n,\rho\nu}\tilde{h}^{\mu\rho}\tilde{h}^{\lambda\nu}\tilde{h}_{\alpha\beta}]\\
\nonumber&\quad -\frac{1}{4}\widetilde{\nabla}_{\alpha}\Phi^{mn}\widetilde{\nabla}_{\beta}\Phi_{mn}
+\frac{1}{8}\Phi^{pq}\widetilde{\nabla}_{\alpha}\Phi_{pq}\Phi^{mn}\widetilde{\nabla}_{\beta}\Phi_{mn}
+\frac{1}{4}
\Phi^{pq}\widetilde{\nabla}_{\mu}\Phi_{mp}\Phi^{mn}\widetilde{\nabla}_{\lambda}\Phi_{nq}\tilde{h}^{\mu\lambda}\tilde{h}_{\alpha\beta}\\
\nonumber&\quad+\frac{3}{8}\tilde h^{\mu\lambda}\widetilde{\nabla}_{\mu}\Phi^{mn}\widetilde{\nabla}_{\lambda}\Phi_{mn}\tilde{h}_{\alpha\beta}-\frac{1}{16}\tilde{h}^{\mu\lambda}
\Phi^{pq}\widetilde{\nabla}_{\mu}\Phi_{pq}\Phi^{mn}\widetilde{\nabla}_{\lambda}\Phi_{mn}\tilde{h}_{\alpha\beta}.
}}

We use in the following the standard 3+1-dimensional ADM formalism for the spacetime metric as in \eqref{ADM-ansatz}, where $\tilde N$, $\tilde g$ and $\tilde X$ are lapse, physical metric and shift vector field. 
The matter quantities appearing in the ADM-Einstein equations in \cite{Re08} read
\eq{\alg{
\tilde \rho&=\tilde N^2 \tilde T^{00},\quad\tilde\jmath_i= \tilde N \tilde T^0_i ,\,\quad \tilde S_{ij}=8\pi(\tilde T_{ij}-\frac12\tilde g_{ij}\tilde g^{kl}\tilde T_{kl})-4\pi \tilde\rho \tilde g_{ij},\quad\tilde \eta=4\pi (\tilde \rho+\tilde g^{ij}\tilde T_{ij}).
}}
Here, $0$ refers to the time-function $\tau$.
\subsection{Rescaled system}
We perform now the rescaling of the macroscopic Einstein equations according to \eqref{rescaling1}.  All symbols in the following denote the rescaled variables as in \eqref{rescaling1}. Then, the Einstein flow in CMCSH gauge reads
\begin{eqnarray}
R(g)-\abg{\Si}^2+\tfrac {2}{3}&=&4\tau\cdot\rho \label{Co-Ha},\\
D^i\Si_{ij}&=&\tau^2 \jmath_j\label{Co-Mo},\\
\left(\De-\tfrac13\right) N&=& N\Big(\abg{\Si}^2+{\tau\cdot \eta}\Big)-1\label{lapse},\\
\De X^i+R^i_{\, j}X^j&=&2D_jN\Si^{ji}-{D^i\left(\tfrac N3-1\right)}_{}+2N \tau^2{\jmath^i}\label{shift}-(2N\Si^{jk}-D^jX^k)(\Chr ijk-\hChr ijk),\\
\p T g_{ij}&=&{2N\Si_{ij}}_{}+2\left(\tfrac{N}3-1\right)g_{ij}-\mcr{L}_Xg_{ij}\label{ev-g},\\
\p T\Si_{ij}&=&{-2\Si_{ij}-N\left(R_{ij}-\delta_{ij}+\tfrac{2}{9}g_{ij}\right)}_{}\label{ev-k}+D^2_{ij}N+2N\Si_{ik}\Si^{k}_j\nonumber\\
&&-\tfrac13\left(\tfrac{N}{3}-1\right)g_{ij}-\left(\tfrac{N}{3}-1\right)\Si_{ij}\nonumber-\mcr{L}_X\Si_{ij}+ {N\tau\cdot S_{ij}}.\nonumber
\end{eqnarray}

The relation between the rescaled matter quantities and the original ones is
\eq{\label{matter-rescaled}
\alg{
\rho&:=4\pi\tilde{\rho}\cdot \tau^{-3}, \quad \eta:=4\pi(\tilde{\rho}+\tilde{g}^{ij}\tilde{T}_{ij})\cdot \tau^{-3},\quad
\underline{\eta}:=4\pi \tilde g^{ij}\tilde T_{ij}\cdot\tau^{-5},\\
\jmath^i&:=8\pi \ab{\tau}^{-5} \tilde \jmath^i, \quad{S}_{ij}:=8\pi{\tau}^{-1}\left[\tilde{T}_{ij}-\tfrac1{2}\tilde{g}_{ij}\tilde{T}\right].
}
} 
This set of equations is the basis for analyzing the dynamical behaviour of the perturbations of the geometry of the macroscopic space.

Before going on we define our notion of smallness. In what follows we say a solution or data is \emph{small} or fulfills a \emph{smallness condition} if \eq{
\Abk{g-\gamma}4+\Abk{\Si}3+\Abk{N-3}5+\Abk{X}5+\Abk{\Phi-\Phi_{\mathsf{b}}}4+\Abk{F}3<\varepsilon
}
for a sufficiently small $\varepsilon>0$, where $\Phi_{\mathsf{b}}$ is a fixed constant map. By construction this condition holds for the initial perturbation we consider, then by local stability of the system, the condition holds on a finite time-interval. This justifies to make the smallness assumption to derive the decay estimates in the sense of a standard bootstrap argument.  
\section{Estimating the energy-momentum tensor} \label{sec : EMT}
In this section we evaluate the matter terms in the Einstein equations in terms of the wave-type-map and the one-forms determining the energy-momentum tensor. We clarify important notations prior to the computations.
The index $0$ corresponds to the $\tau$ time-function in this section and we use the notation $\tr$ to compute the \emph{trace of the Lie-algebra indices}.  We first compute the rescaled energy density
\eq{\label{en-den-exp}\alg{
\rho&=4\pi \tau^{-3} \tilde N^2 \tilde T^{00}=4\pi\tau^{-3}\tilde N^2\tilde h^{0\mu}\tilde h^{0\nu}\tilde T_{\mu\nu}
=4\pi\tau N^{-2}(\tilde T_{00}+ \tau^{-2}X^i X^j\tilde T_{ij}).
}}
We evaluate $T_{00}$ in the following. 
\eq{\label{ex-t00}\alg{
\tilde T_{00}=&\frac{1}{2}\sqrt{\det\Phi}\tr(F^\mu_{~0}F_{0\mu})-\frac{1}{4}\tilde{\nabla}_0\Phi^{mq}\tilde{\nabla}_0\Phi_{mq} 
+\frac{1}{8}(\Phi^{mq}\tilde{\nabla}_0\Phi_{mq})(\Phi^{mq}\tilde{\nabla}_0\Phi_{mq})\\
&+\Bigg[\big(-\frac{1}{4}\sqrt{\det\Phi}\tr(F^\mu_{~\gamma}F_{\mu}^\gamma)+\frac{3}{8}\tilde{\na}^\gamma\Phi^{mq}\tilde{\na}_\gamma\Phi_{mq}
-\frac{1}{16}(\Phi^{mq}\tilde{\na}_\gamma\Phi_{mq})(\Phi^{mq}\tilde{\na}^\gamma\Phi_{mq})\big)\\
&\qquad+\Phi^{mn}(\frac14\Phi^{pq}\tilde{\na}_\lambda\Phi_{mp}\tilde{\na}^\lambda\Phi_{nq}+\frac18\sqrt{\det\Phi}(F_{\mu\nu,m}F^{\mu\nu}_n))\Bigg]\tau^{-4}(-N^2+\abg{X}^2)
}}
The spatial part is given by
\eq{\alg{
\tilde T_{ij}=&\frac{1}{2}\sqrt{\det\Phi}\tr(F^\mu_{~j}F_{i\mu})-\frac{1}{4}\tilde{\nabla}_i\Phi^{mq}\tilde{\nabla}_j\Phi_{mq} 
+\frac{1}{8}(\Phi^{mq}\tilde{\nabla}_i\Phi_{mq})(\Phi^{mq}\tilde{\nabla}_j\Phi_{mq})\\
&+\Bigg[\big(-\frac{1}{4}\sqrt{\det\Phi}\tr(F^\mu_{~\gamma}F_{\mu}^\gamma)+\frac{3}{8}\tilde{\na}^\gamma\Phi^{mq}\tilde{\na}_\gamma\Phi_{mq}
-\frac{1}{16}(\Phi^{mq}\tilde{\na}_\gamma\Phi_{mq})(\Phi^{mq}\tilde{\na}^\gamma\Phi_{mq})\big)\\
&\qquad+\Phi^{mn}(\frac14\Phi^{pq}\tilde{\na}_\lambda\Phi_{mp}\tilde{\na}^\lambda\Phi_{nq}+\frac18\sqrt{\det\Phi}(F_{\mu\nu,m}F^{\mu\nu}_n))\Bigg]\tau^{-2} g_{ij}.
}}

We evaluate now the trace-part of $\eta$
\eq{\label{s-trace}\alg{
\tilde g^{ij}\tilde{T}_{ij}=&\tau^2g^{ij}\Big[\frac{1}{2}\sqrt{\det\Phi}\tr(F^\mu_{~j}F_{i\mu})-\frac{1}{4}\tilde{\nabla}_i\Phi^{mq}\tilde{\nabla}_j\Phi_{mq} 
+\frac{1}{8}(\Phi^{mq}\tilde{\nabla}_i\Phi_{mq})(\Phi^{mq}\tilde{\nabla}_j\Phi_{mq})\Big]\\
&+3\Bigg[\big(-\frac{1}{4}\sqrt{\det\Phi}\tr(F^\mu_{~\gamma}F_{\mu}^\gamma)+\frac{3}{8}\tilde{\na}^\gamma\Phi^{mq}\tilde{\na}_\gamma\Phi_{mq}
-\frac{1}{16}(\Phi^{mq}\tilde{\na}_\gamma\Phi_{mq})(\Phi^{mq}\tilde{\na}^\gamma\Phi_{mq})\big)\\
&\qquad+\Phi^{mn}(\frac14\Phi^{pq}\tilde{\na}_\lambda\Phi_{mp}\tilde{\na}^\lambda\Phi_{nq}+\frac18\sqrt{\det\Phi}(F_{\mu\nu,m}F^{\mu\nu}_n))\Bigg].
}}

We evaluate the current
\eq{\label{j-exp}\alg{
\jmath^j&=8\pi \ab{\tau}^{-5} \tilde N \tilde T_i^0\tilde g^{ij}=8\pi \tau^{-2}N^{-1}g^{ij}(-\tilde T_{0i}\ab{\tau}+\tilde T_{ki}X^k).
}}
Here, we require the off-diagonal components of the energy-momentum tensor. Those are given below

\eq{\label{off-diag}\alg{
\tilde T_{0 i}&=\frac{1}{2}\sqrt{\det\Phi}\tr(F^\mu_{~0}F_{i\mu})-\frac{1}{4}\tilde{\nabla}_0\Phi^{mq}\tilde{\nabla}_i\Phi_{mq} 
+\frac{1}{8}(\Phi^{mq}\tilde{\nabla}_0\Phi_{mq})(\Phi^{mq}\tilde{\nabla}_i\Phi_{mq})\\
&\quad+\Bigg[\big(-\frac{1}{4}\sqrt{\det\Phi}\tr(F^\mu_{~\gamma}F_{\mu}^\gamma)+\frac{3}{8}\tilde{\na}^\gamma\Phi^{mq}\tilde{\na}_\gamma\Phi_{mq}
-\frac{1}{16}(\Phi^{mq}\tilde{\na}_\gamma\Phi_{mq})(\Phi^{mq}\tilde{\na}^\gamma\Phi_{mq})\big)\\
&\qquad+\Phi^{mn}(\frac14\Phi^{pq}\tilde{\na}_\lambda\Phi_{mp}\tilde{\na}^\lambda\Phi_{nq}+\frac18\sqrt{\det\Phi}(F_{\mu\nu,m}F^{\mu\nu}_n))\Bigg]\ab\tau^{-3}X_i.
}}

\subsection{$L^2$-norms of the Energy-Momentum tensor}
When analyzing the dynamics of the metric variables we require bounds on standard Sobolev norms of the matter variables as listed in \eqref{matter-rescaled}. Those correspond directly to bounds on the Sobolev norms of components of the energy-momentum tensor. We derive those bounds in the following, expressed in terms of the corresponding norms of the matter fields $F$, $A$ and $\Phi$, respectively. We define some useful norms for this purpose.
\eq{\alg{
\Abbk{F}\ell^2&:= \sum_{m}\sum_{k\leq \ell}\int _M \Big(\ab{\tau}^{2}g^{ij} D_{i_1}\hdots D_{i_k} (F_{0i,m})D^{i_1}\hdots D^{i_k} (F_{0j,m})\\
&\qquad\qquad\qquad+g^{ij}g^{uv} D_{i_1}\hdots D_{i_k} (F_{iu,m})D^{i_1}\hdots D^{i_k} (F_{jv,m})\Big)dV_g
}}
Moreover, $F$ is antisymmetric and the factor $\ab{\tau}^{2}$ compensates a growth of the $0$-components of $F$ relative to the pure spatial components. It needs to be determined/fixed as soon as the decay properties of $F$ are understood. Note that all objects and derivatives here are defined with respect to the rescaled metric $g$ such that there is no more intrinsic scaling in this energy. More precisely, this means that under the condition that the rescaled metric remains close to the reference metric $\gamma$, this energy measures the field $F$ without introducing a growth resulting from the expansion as it would be the case for the unrescaled physical metric.

\begin{rem}
Note that the tensor $F$ is given in terms of derivatives of the 1-form $A$ here with respect to the dual basis $(d\tau,dx^1,dx^2,dx^3)$. As shown in the analysis of the asymptotic behavior of $A$, the coefficients of $A$ are controlled when expressed with respect to the basis $(dT,dx^1,dx^2,dx^3)$. From the comparison of the bases we obtain that the coefficients are related via $A_\tau=\frac{dT}{d\tau} A_T=-\tau^{-1}A_T$.
\end{rem}

Analogously, we define for the field $\Phi$ a similar Sobolev norm.

\eq{\alg{
\Abbk{\Phi}\ell^2&:=\sum_{m,n}\Bigg[ \sum_{k\leq \ell-1}\int_M \left((D_{i_1}\hdots D_{i_k}(\partial_0 \Phi_{mn}) D^{i_1}\hdots D^{i_k}(\partial_0 \Phi_{mn})     \right)dV_g\\
&\qquad\qquad +\sum_{k\leq \ell}\int_M \tau^{-2}\left(( D_{i_1}\hdots D_{i_k}( \Phi_{mn})  D^{i_1}\hdots D^{i_k}( \Phi_{mn})\right)dV_g\Bigg]
}}

We obtain the following estimates for the components of the energy-momentum tensor as appearing in the matter variables.
Recall, $\widehat X=X/N$. Then we find the following
\begin{lem}
Let $\ell\geq3/2$. Then the following estimate holds
\eq{\alg{
\Abk{\tilde T_{00}}\ell+\ab{\tau}^{-1}\Abk{\tilde T_{0i}}\ell +\ab{\tau}^{-2}\Abk{\underline {\tilde T}}\ell+\ab{\tau}^{-4}\Abk{\tilde g^{ij}\tilde T_{ij}}\ell&\leq C(1+\Abk{\widehat X}{\ell}^2)\Big[  \Abbk{F}\ell^2 +\Abbk{\Phi}\ell^2\Big].
}}
Here, $C=C(\Abi{N},\Abi{N^{-1}},\Abi{\Phi},\Abi{\Phi^{-1}},\Abk{N}\ell)$ and we denote by $\underline {\tilde T}$ the spatial part of the tensor $\tilde T$.
\end{lem}
\begin{proof}
The estimates follow immediately from the expansions of the energy-momentum tensor components \eqref{ex-t00}-\eqref{s-trace} and \eqref{off-diag}.
\end{proof}

\begin{rem}
Note that the constant, given that all arguments are uniformly bounded, as we assure by suitable bootstrap assumptions, can be considered a generic constant.
\end{rem}

\begin{proof}
We evaluate first the term, $\tilde T_{00}$. For the first estimate we evaluate 
\eq{\alg{
\mathrm{tr} \left(F^\mu_0F_{\mu 0}\right)=\tau^2(g^{ij}-\widehat X^i\widehat X^j)\delta ^{mn}F_{i0,m}F_{0j,n}
}}
and then apply Sobolov embedding with $\ell>3/2$. We evaluate next the square of $F$, which is
\eq{\alg{
\delta ^{mn}F_{\gamma, m}^\mu F^\gamma_{\mu, n}=2\delta ^{mn}\Big[-\tau^6 N^{-2} g^{ij}F_{i0,m}F_{j0,n}+2 \tau^6 N^{-2}\hat X^i\hat X^j F_{0i,m}F_{jb0,n}\\
+\tau^5 (g^{ij}-\hat X^i\hat X^j)N^{-1}\hat X^v\left(F_{iv,m}F_{j0,n}+F_{iv,n}F_{j0,m}\right)\\
 +\frac12 \tau^4 (g^{uv}-\hat X^u\hat X^v)(g^{ij}-\hat X^i\hat X^j)F_{iv,m}F_{uj,n}\Big].
}}
The related term containing a square of $F$ , where $\delta ^{mn}$ is replaced by $\Phi^{mn}$ can be decomposed identically by replacing $\delta $ by $\Phi$. For example, we consider the following term
\eq{\alg{
\tilde\nabla^\gamma \Phi^{mq} \tilde\nabla_\gamma \Phi_{mq}&= -\tau^4\frac1{N^2} \partial_0 \Phi^{mq} \partial_0 \Phi_{mq}+\tau^3 \frac1N \hat X^i (\partial_0 \Phi^{mq} \partial_i\Phi_{mq}+\partial_i\Phi ^{mq}\partial_0 \Phi_{mq})\\
&\quad+\tau^2 (g^{ij}-\hat X^i\hat X^j)\partial_i\Phi^{mq}\partial_j \Phi_{mq}.
}}
Similar decompositions hold for the other terms in the brackets on the right-hand side of the equation for $\tilde T_{00}$. The remaining terms in the first line of that equation can be estimated directly. To deduce the full estimate it is sufficient to use the fact that the regularity is high enough to use product estimates for the Sobolev norm and that every time derivative of $\Phi$ and every zero-component of $F$ appears with one additional $\tau$ factor.\\
We turn now to the estimate for $\tilde T_{0i}$. We note that the term in the big brackets is identical to the case considered above. Since the last factor is now only a $\tau^{-3}$ and a shift term we obtain the first summand in the estimate. 
It remains to evaluate the first line of the evaluation of $\tilde T_{0i}$. The terms containing derivatives of $\Phi$ can immediately be estimated. We evaluate the first term containing $F$.
\eq{
F^\mu_{0,m}F_{i\mu,n}=\tau^3 \frac{\hat X^v}{N}F_{v0,m}F_{i0,n}+\tau^2(g^{uv}-\hat X^u\hat X^v)F_{v0,m}F_{iu,n}
}
These terms yield terms that decay like $\tau$ as contained in the estimate. We turn to the estimate for $\underline {\tilde T}$ now noting that the trace of $\underline {\tilde T}$ can be treated similarly. The term in the large brackets is unchanged and is here multiplied only with a $\tau^{-2}$ factor. This leaves an overall $\tau^2$ factor, which appears in the estimate.
\end{proof}

\subsection{Estimating the matter variables as appearing in the Einstein equations}
The final estimates for the rescaled matter quantities as appearing in the Einstein equations are the following:
\begin{prop}\label{matt-quant-est} Let $\ell\geq 3/2$, then we have
\eq{\alg{
\Abk{\rho}\ell &\leq  C\ab{\tau}(1+\Abk{\hat X}\ell^2)\left[\Abbk{F}\ell^2 +\Abbk{\Phi}\ell^2\right],\qquad \Abk{\eta}\ell \leq C\ab{\tau}(1+\Abk{\hat X}\ell^2)\left[\Abbk{F}\ell^2 +\Abbk{\Phi}\ell^2\right],\\
\Abk{\jmath}\ell&\leq C(1+\Abk{\hat X}\ell^2)\left[\Abbk{F}\ell^2 +\Abbk{\Phi}\ell^2\right],\qquad\,\,\,\,\,\, \Abk{S}\ell \leq C\ab{\tau}(1+\Abk{\hat X}\ell^2)\left[\Abbk{F}\ell^2 +\Abbk{\Phi}\ell^2\right].
}}
\end{prop}
\begin{proof}
This is an immediate consequence of the foregoing lemma as well as \eqref{en-den-exp} and \eqref{j-exp} and the definitions of $\eta$ and $S$.
\end{proof}

\section{Energy estimates} \label{sec : hyp}

\subsection{Energy estimate for the geometry}
We define the energy to measure the tracefree part of the second fundamental form and of the difference between the metric and the background metric as in the related work \cite{AnMo11}.
We recall briefly some necessary notation. The lowest eigenvalue of the Einstein operator corresponding to the specific Einstein metric is denoted by $\la_0$. For a relevant lower bound in the present case cf.~\cite{Kr15}. The correction constants $\al=\al(\la_0,\delta_\alpha)$ and $c_E$ are given by
\eq{
\al=
\begin{cases}
1& \la_0>1/9\\
1-\delta_\alpha& \la_0=1/9
\end{cases}\quad,
\qquad c_E=\begin{cases}
1& \la_0>1/9\\
9(\la_0-\varepsilon')& \la_0=1/9
\end{cases}
}
with $\delta_\alpha=\sqrt{1-9(\la_0-\varepsilon')}$, where $1>>\varepsilon'>0$ is a free variable to be chosen below. 
The energy is defined in the following. For $m\geq 1$ let

\eq{\alg{
\mathcal{E}_{(m)}=\frac12\int_M\langle 6\Si,\mcl L_{g,\ga}^{m-1} 6\Si\rangle dV_g+\frac92\int_M \langle (g-\gamma),\mcl L_{g,\ga}^{m}(g-\gamma)\rangle dV_g,\quad
\Gamma_{(m)}=\int_M \langle 6\Si,\mcl L_{g,\ga}^{m-1}(g-\gamma)\rangle dV_g.
}}

The corrected energy is 
\eq{
E_s(g-\gamma,\Si)=\sum_{1\leq m\leq s} \mcl E_{(m)}+c_E\Gamma_{(m)}.
}

\begin{lem}
There exists a $\delta>0$ and a constant $C>0$ such that for $\delta$-small data $(g,\Si,A,\Phi)$ the inequality
\eq{
\Abk{g-\gamma}s^2+\Abk{\Si}{s-1}^2\leq C E_s(g,\Si)
}
holds.
\end{lem}

\begin{proof}
This is analogous to the previous work \cite{AnMo11} taking into account the triviality of the kernel of the Einstein operator. 
\end{proof}

The relevant energy estimate for the corrected energy is
\begin{lem}\label{lem-geom-en-est}
For sufficiently small $E_s$ we have
\eq{\label{en-est-geom}\alg{
\p T E_s &\leq -2\al E_s+6 E_s^{1/2}\ab{\tau}\Abk{NS}{s-1}+ CE_s^{3/2} +C E_s^{1/2}\left(\ab{\tau}\Abk{\rho}{s-1}+\ab{\tau}^3\Abk{\underline\eta}{s-1}+\ab{\tau}^2\Abk{N\jmath}{s-2}\right).
}}
\end{lem}

\begin{proof}
The proof is analogous to the one in \cite{AnFa17}.
\end{proof}

Substituting the norms of the matter quantities by Proposition \ref{matt-quant-est} we obtain the energy estimate. 

\begin{prop}
Let $s\geq 5/2$ and $E_s$ be sufficiently small. Then we have
\eq{\label{en-est-geom-impr}\alg{
\p T E_s &\leq -2\al E_s+C E_s^{1/2}\ab{\tau}^2\left[\Abbk{F}{s-1}^2 +\Abbk{\Phi}{s-1}^2\right]+ CE_s^{3/2} ,
}}
where $C=C(\Abk{\hat X}{s-1},\Abi{N},\Abi{N^{-1}},\Abk{N}{s-1})$.
\end{prop}

\subsection{Energy estimates for the vector potential}\label{sec : 62}
With respect to the Lorentzian metric $h$ defined in \eqref{metric-h}, \eqref{maxwell-eq} can be written as
\begin{align}\label{curv-eq-3}
{h}^{\lambda\mu}{\nabla}_{\lambda}F_{\mu\alpha}=-\frac{1}{2}{h}^{\lambda\mu}F_{\lambda\alpha}\Phi^{mp}\partial_{\mu}\Phi_{mp}.
\end{align}
Here and in the rest of the subsection, we omit the index $l$ in equation \eqref{maxwell-eq} due to convenience.

\begin{lem}[Slice-adapted gauge]\label{gauge}
Let $F\in \Omega^2(\widetilde{M})$ be exact. 
Then there exists a unique form $A\in \Omega^1(\widetilde{M})$ with $dA=F$ such that
\begin{align}
\mathrm{div}_g\omega=0 ,\qquad
\qquad \omega\perp\ker(\Delta_{H}), \qquad\int_M \Psi dV_g=0,
\end{align}
where $\omega$ and $\Psi$ are defined in Section \ref{formalism}.
In the statement and the proof of the lemma, $d$ is the exterior derivative on $\widetilde{M}$.
\end{lem}
\begin{proof}	
Let $B\in \Omega^1(\widetilde{M})$ such that $dB=F$. Let $f\in C^{\infty}(\widetilde{M})$ with $\int_MfdV_g=0$ for each $T\in I$, $c\in C^{\infty}(I)$ and $\eta \in C^{\infty}(I,\Omega^1(M))$ be such that $\eta\in \ker(\Delta_{H})$ for each $T\in I$.
Let 
\begin{align}A=B+d(f+c)-\eta\in \Omega^1(\widetilde{M}).
\end{align}
By construction, $dA=dB$. Demanding the first condition of the lemma yields $g^{ij}D_iB_j+\Delta_gf=0$
and because $\int_M g^{ij}D_iB_jdV_g=0$, this equation can be uniquely solved at each time. Let $\omega_1,\ldots ,\omega_L\in C^{\infty}(I,\Omega^1(M)) $ be for each $T$ an $L^2(g)$-orthonormal basis of $\ker(\Delta_{H})$ (Note that the dimension of $\ker(\Delta_{H})$ equals the first Betti number of $M$. Thus, it does not depend on $g$). The second gauge condition is obtained by defining
\begin{align}
\eta=\sum_{a=1}^L\int_M \langle B+d(f+c),\omega_a\rangle_g dV_g\cdot \omega_a=\sum_{a=1}^L\int_M \langle B,\omega_a\rangle_g dV_g\cdot \omega_a.
\end{align}
The third condition yields
\begin{align}
\int_{M}A(e_0)dV_g=\int_M(B(e_0)+df(e_0)dV_g+\partial_Tc\cdot\int_MN^{-1}dV_g,
\end{align}
which fixes $\partial_Tc$. Uniqueness of $A$ follows by construction.
\end{proof}
\begin{prop}\label{form-eq-1}Let $F\in \Omega^2(\widetilde{M})$ be exact and assume it solves \eqref{curv-eq-3}. Let $A\in\Omega^1(\widetilde{M})$ be a potential for $F$ which satisfies the gauge conditions of Lemma \ref{gauge} and let $\Psi$ and $\omega$ be as in Lemma \ref{gauge}. Then we have the equations
\begin{align}\label{form-eq-11}
\Delta_g(\Psi)&=-\mathrm{div}_g(\Psi\cdot d(\log(N)))-[\mathcal{L}_{e_0},\mathrm{div}_g]\omega-\frac{1}{2}g^{ij}F_{i 0}\Phi^{mp}\partial_{j}\Phi_{mp},\\
\begin{split}\label{form-eq-12}
(\mathcal{L}_{e_0}(\mathcal{L}_{e_0}\omega))_k+\Delta_H\omega_k&=\partial_k(\partial_{e_0}\Psi)+\partial_{e_0}\Psi\frac{\partial_kN}{N}+\Psi\cdot \partial_k(\partial_{e_0}\log(N))+g^{ij}\frac{\partial_iN}{N}F_{j k}-g^{ij}\Pi_{ki}F_{0j}\\
&\qquad+\tr_g\Pi\cdot F_{0k}+g^{ij}\Pi_{ik}F_{j0}-\frac{1}{2}F_{0 k}\Phi^{mp}\partial_{0}\Phi_{mp}
+\frac{1}{2}h^{ij}F_{i k}\Phi^{mp}\partial_{j}\Phi_{mp}.
\end{split}
\end{align}
\end{prop}
\begin{proof}
Using \eqref{christoffel}, one computes
\begin{align}
g^{ij}{\nabla}_iF_{j0}
&=\div_g(i_{e_0}F),
\end{align}
where $i_{e_0}F\in  C^{\infty}(I,\Omega^1(M))$ is given by $i_{e_0}F(\partial_i)=F(\partial_i,e_0)$.
Moreover, by using $\omega(e_0)=0$, $A(\partial_i)=\omega(\partial_i)$ and $[\partial_i,e_0]=-\frac{\partial_iN}{N}e_0+N^{-1}[\partial_i,X]$, we find
\begin{align}\label{i_{e_0}F}
i_{e_0}F(\partial_i)
=\partial_i(\Psi)+\frac{\partial_iN}{N}\Psi-\mathcal{L}_{e_0}\omega(\partial_i)
\end{align}
which implies
\begin{align}
g^{ij}{\nabla}_iF_{j0}=\div_g(i_{e_0}F)
=\Delta_g\Psi+\div_g(\Psi\cdot D(\log(N)))+[\mathcal{L}_{e_0},\div]\omega,
\end{align}
where we have used that $\div_g\omega=0$. The first formula follows from \eqref{curv-eq-3}.
To prove the second formula, we compute, using \eqref{christoffel} again,
\begin{align}
g^{ij}{\nabla}_iF_{jk}
=-\Delta_H\omega_k+\tr_g\Pi\cdot F_{0k}+g^{ij}\Pi_{ik}F_{j0}
\end{align}
and
\begin{align}
\nabla_0F_{0k}
=(\mathcal{L}_{e_0}(\mathcal{L}_{e_0}\omega))_k-\partial_k(\partial_{e_0}\Psi)-\partial_{e_0}\psi\frac{\partial_kN}{N}-\Psi\cdot \partial_k(\partial_{e_0}\log(N))-g^{ij}\frac{\partial_iN}{N}F_{j k}
+g^{ij}\Pi_{ki}F_{0j}.
\end{align}
Therefore, the second formula again follows from \eqref{curv-eq-3}.
\end{proof}
\begin{rem}
Local existence for the system \eqref{form-eq-11},\eqref{form-eq-12} is argued as follows: One first solves \eqref{curv-eq-3} by using the Lorentz gauge $h^{\lambda\mu}\nabla_{\lambda}A_{\mu}=0$. In this gauge, \eqref{form-eq-1} becomes
\begin{align}\label{form-eq-2}
\Box_{H,h}A_\alpha=\frac{1}{2}{h}^{\lambda\mu}F_{\lambda\alpha}\Phi^{mp}\partial_{\mu}\Phi_{mp}
\end{align}
 for which local existence follows from standard theory. Here, $\Box_{H,h}$ denotes the Hodge wave operator of the metric $h$.
 By the construction in the proof of Lemma \ref{gauge}, we obtain in a unique way a pair $(\omega,\Psi)$ which solves the system \eqref{form-eq-11},\eqref{form-eq-12}. On the other hand, as long as the solution $(\omega,\Psi)$ of \eqref{form-eq-11},\eqref{form-eq-12} is bounded, any corresponding solution $A$ of \eqref{form-eq-2} is also bounded: Let $B=\omega+\Psi\cdot e_0^*$ and $f$ be a solution of the equation
\begin{align}
\Box_hf=h^{\lambda\mu}\nabla_{\lambda}B_{\mu}.
\end{align}
Then $A=B+df$ satisfies the Lorentz gauge and is bounded by construction. The main advantage of the slice-adapted gauge is that it is easier to control the solution \eqref{form-eq-11},\eqref{form-eq-12} by energy estimates.
\end{rem}

The structure of the system \eqref{form-eq-11},\eqref{form-eq-12} motivates the following energy
\begin{align}\label{energy-vector-potential}
E_k(\omega)=\sum_{l=0}^{k-1}\int_M( \langle (\Delta_H)^l\mathcal{L}_{e_0}\omega,\mathcal{L}_{e_0}\omega\rangle_g+\langle (\Delta_H)^{l+1}\omega,\omega\rangle_g)dV\simeq \left\|\mathcal{L}_{e_0}\omega\right\|_{H^{k-1}(g)}^2+ \left\|\omega\right\|_{H^{k}(g)}^2
\end{align}
with $k\geq1$.
Note that due to the gauge condition $\omega\perp\ker(\Delta_H)$ and elliptic regularity, the $L^2$-norm of $\omega$ is controlled by $E_k(\omega)$.
\begin{lem}
Suppose that $F$ solves \eqref{curv-eq-3}, $A\in \Omega^1(\widetilde{M})$ is a gauged vector potential for $F$ and $\Psi$ and $\omega$ are as in Lemma \ref{gauge}. Then for $k> n/2+1$ and provided that $N$ is uniformly positive and $\left\| N\right\|_{H^{k}}$ is bounded by some fixed constant, we have the energy estimate
\begin{equation}
\begin{split}
\partial_TE_k&\leq
C(\left\|D\log N\right\|_{H^{k-1}}+\left\|\Pi\right\|_{H^{k-1}}+\left\|S\right\|_{H^{k-1}}
+\left\|D\Phi\right\|_{H^{k-1}}+\left\|\partial_{e_0}\Phi\right\|_{H^{k-1}})E_k\\
&\quad+C(\left\|\partial_{e_0}\Psi\right\|_{H^{k}}+(
\left\|\partial_{e_0}\log N\right\|_{H^{k}}+\left\|\Pi\right\|_{H^{k-1}}+\left\|\partial_{e_0}\Phi\right\|_{H^{k-1}}
)\left\|\Psi\right\|_{H^{k}})\sqrt{E_k}.
\end{split}
\end{equation}
\end{lem}
\begin{proof}
First recall that the Hodge Laplacian is defined as $\Delta_H=dd^*+d^*d$. By extending this definition to the exterior algebra $\Omega^*(M)$, we may also write $\Delta_H=(d+d^*)^2$ where $d+d^*$ is a self-adjoint first-order differential operator acting on the exterior algebra. Fix $l\in\left\{0,\ldots,k-1\right\}$. By integration by parts,
\begin{equation}
\begin{split}
\int_M (\langle &(\Delta_H)^l\mathcal{L}_{e_0}\omega,\mathcal{L}_{e_0}\omega\rangle_g+\langle (\Delta_H)^{l+1}\omega,\omega\rangle_g)dV\\
&=\int_M (\langle (d+d^*)^l\mathcal{L}_{e_0}\omega,(d+d^*)^l\mathcal{L}_{e_0}\omega\rangle_g+\langle (d+d^*)^{l+1}\omega,(d+d^*)^{l+1}\omega\rangle_g)dV=:\int_M\mathcal{E}_l dV.
\end{split}
\end{equation}
At first, we compute
\begin{align}
\partial_T\int_M\mathcal{E}_ldV
&=\int_M N(\partial_{e_0}\mathcal{E}_l)dV-\int_MN\cdot \mathcal{E}_l\tr_g\Pi dV.
\end{align}
In the following we will make use of the \(*\)-notation to denote various contractions between tensors.
Therefore, after integration by parts we get
\begin{align}\begin{split}\label{bigeq}
\partial_T\int_M\mathcal{E}_ldV&=\int_M N\cdot\Pi*((d+d^*)^l\mathcal{L}_{e_0}\omega)*((d+d^*)^l\mathcal{L}_{e_0}\omega )dV\\
&\quad+\int_M N\cdot\Pi*((d+d^*)^{l+1}\omega)*((d+d^*)^{l+1}\omega )dV\\
&\quad+2\int_M N\cdot \langle [\mathcal{L}_{e_0},(d+d^*)^l]\mathcal{L}_{e_0}\omega,(d+d^*)^l\mathcal{L}_{e_0}\omega\rangle dV\\
&\quad+2\int_M N\cdot \langle [\mathcal{L}_{e_0},(d+d^*)^{l+1}]\omega,(d+d^*)^{l+1}\omega\rangle dV\\
&\quad + \int_M D N*(d+d^*)^l(\mathcal{L}_{e_0}\omega)*((d+d^*)^{l+1}\omega) dV\\
&\quad +\int_M \langle (d+d^*)^l(\mathcal{L}_{e_0}(\mathcal{L}_{e_0}\omega)+\Delta_H\omega),(d+d^*)^l\mathcal{L}_{e_0}\omega\rangle dV
-\int_MN\cdot \mathcal{E}_l\tr_g\Pi dV.
\end{split}
\end{align}
We have 
\begin{equation}
\begin{split}
&\int_M N\cdot\Pi*((d+d^*)^l\mathcal{L}_{e_0}\omega)*((d+d^*)^l\mathcal{L}_{e_0}\omega )dV\\
&\quad+\int_M N\cdot\Pi*((d+d^*)^{l+1}\omega)*((d+d^*)^{l+1}\omega )dV-\int_MN\cdot \mathcal{E}_l\tr_g\Pi dV
\leq C\left\| N\right\|_{H^{k-2}}\left\| \Pi\right\|_{H^{k-2}}E_k,
\end{split}
\end{equation}
and 
\begin{align}
\int_M D N*(d+d^*)^l(\mathcal{L}_{e_0}\omega)*((d+d^*)^{l+1}\omega) dV
&\leq C\left\|D N\right\|_{H^{k-2}}E_k.
\end{align}
Now we estimate the commutator terms. Similarly as in Lemma \ref{variations}, we have
\begin{align}
[\mathcal{L}_{e_0},(d+d^*)]\eta=[\mathcal{L}_{e_0},d^*]\eta =\Pi*D\eta+D\log N*\mathcal{L}_{e_0}\eta+ S*\eta+\Pi *D\log N*\eta
\end{align}
for a general differential form $\eta\in C^{\infty}(I,\Omega^m(M))$.
Here, we used the notation $S=2\div_g\Pi-D\tr_g\Pi$.
By induction, we get
\begin{equation}
\begin{split}
[\mathcal{L}_{e_0},(d+d^*)^{l}]\mathcal{L}_{e_0}\omega&=\sum_{m=0}^{l-1}D^m\Pi*D^{l-m}\mathcal{L}_{e_0}\omega
+\sum_{m=0}^{l-1}D^{m+1}\log N*D^{l-1-m}(\mathcal{L}_{e_0}(\mathcal{L}_{e_0}\omega))\\
&\quad +\sum_{m=0}^{l-1}D^{m+1}\log N*[\mathcal{L}_{e_0},(d+d^*)^{l-1-m}]\mathcal{L}_{e_0}\omega+\sum_{m=0}^{l-1} D^{m} S*D^{l-1-m}\mathcal{L}_{e_0}\omega 
\\&\quad +\sum_{m=0}^{l-1} D^{m}(\Pi*D\log N)*D^{l-1-m}\mathcal{L}_{e_0}\omega,
\end{split}
\end{equation}
and again by induction,
\begin{equation}
\begin{split}
[\mathcal{L}_{e_0},(d+d^*)^{l}]\mathcal{L}_{e_0}\omega&=
\sum_{n=0}^{l-1}\sum_{\sum l_i+p=n}\sum_{p=0}^n\underbrace{D^{l_1+1}\log N*\ldots D^{l_p+1}\log N}_{p-\text{times}}*\\
&\quad [\sum_{m=0}^{l-1-n}D^m\Pi*D^{l-n-m}\mathcal{L}_{e_0}\omega	+\sum_{m=0}^{l-1-n}D^{m+1}\log N*D^{l-1-n-m}(\mathcal{L}_{e_0}(\mathcal{L}_{e_0}\omega))\\
&\quad+\sum_{m=0}^{l-1-n} D^{m} S*D^{l-1-n-m}\mathcal{L}_{e_0}\omega 
+\sum_{m=0}^{l-1-n} D^{m}(\Pi*D\log N)*D^{l-1-n-m}\mathcal{L}_{e_0}\omega
].
\end{split}
\end{equation}
Similarly, we get
\begin{equation}
\begin{split}
[\mathcal{L}_{e_0},(d+d^*)^{l+1}]\omega&=
\sum_{n=0}^{l}\sum_{\sum l_i+p=n}\sum_{p=0}^n\underbrace{D^{l_1+1}\log N*\ldots D^{l_p+1}\log N}_{p-\text{times}}*\\
&\quad [\sum_{m=0}^{l-n}D^m\Pi*D^{l+1-n-m}\omega	+\sum_{m=0}^{l-n}D^{m+1}\log N*D^{l-n-m}\mathcal{L}_{e_0}\omega\\
&\quad+\sum_{m=0}^{l-n} D^{m} S*D^{l-n-m}\omega 
+\sum_{m=0}^{l-n} D^{m}(\Pi*D\log N)*D^{l-n-m}\omega
].
\end{split}
\end{equation}	
Therefore, by the bounds on $N$, 
\begin{equation}
\begin{split}
&\int_M N\cdot \langle [\mathcal{L}_{e_0},(d+d^*)^l]\mathcal{L}_{e_0}\omega,(d+d^*)^l\mathcal{L}_{e_0}\omega\rangle dV\\ 
&\leq C
[(\left\|\Pi\right\|_{H^{k-2}}+\left\|S\right\|_{H^{k-2}}+\left\|D\log N\right\|_{H^{k-2}}\left\|\Pi\right\|_{H^{k-2}})E_k
+\left\|D\log N\right\|_{H^{k-2}}\left\|\mathcal{L}_{e_0}(\mathcal{L}_{e_0}\omega)\right\|_{H^{k-2}}\sqrt{E_k}
].
\end{split}
\end{equation}
Similarly, the second commutator term is estimated as
\begin{equation}
\begin{split}
\int_M &N\cdot \langle [\mathcal{L}_{e_0},(d+d^*)^{l+1}]\omega,(d+d^*)^{l+1}\omega\rangle dV\\ &\leq C
[\left\|\Pi\right\|_{H^{k-1}}+\left\|S\right\|_{H^{k-1}}+\left\|D\log N\right\|_{H^{k-1}}(1+\left\|\Pi\right\|_{H^{k-1}})]E_k.
\end{split}
\end{equation}    
Finally, the second last term in \eqref{bigeq}
can be treated by \eqref{form-eq-12} and standard estimates.
\end{proof}
\begin{lem}\label{lem psi}
As long as $\left\|D\log N\right\|_{H^k}+\left\|D\Phi\right\|_{H^{k-1}}$ is small enough, the function $\Psi$ satisfies the estimate
\begin{align}
\left\|\Psi\right\|_{H^{k+1}}\leq
C(\left\|\Pi\right\|_{H^{k-1}}+(1+\left\|\Pi\right\|_{H^{k-1}})\left\|D\log N\right\|_{H^{k-1}}+\left\|S\right\|_{H^{k-1}}
+\left\|\Phi^{-1}\right\|_{H^{k-1}}\left\|D\Phi\right\|_{H^{k-1}})\sqrt{E_k}.
\end{align}
\end{lem} 
\begin{proof}
By elliptic regularity and the first equation in Lemma \ref{form-eq-1},
\begin{align}
\left\|\Psi\right\|_{H^{k+1}}\leq C\cdot 	\left\|\Psi\right\|_{H^{k}}	\left\|D\log N\right\|_{H^{k}}+	\left\|[\mathcal{L}_{e_0},\mathrm{div}_g]\omega\right\|_{H^{k-1}}+\left\|i_{e_0}F\right\|_{H^{k-1}}\left\|\Phi^{-1}\right\|_{H^{k-1}}\left\|D\Phi\right\|_{H^{k-1}}.
\end{align}
By Lemma \ref{variations}, we have
\begin{align}
\left\|[\mathcal{L}_{e_0},\mathrm{div}_g]\omega\right\|_{H^{k-1}}\leq 
C(\left\|\Pi\right\|_{H^{k-1}}+(1+\left\|\Pi\right\|_{H^{k-1}})\left\|D\log N\right\|_{H^{k-1}}+\left\|S\right\|_{H^{k-1}}
)\sqrt{E_k}	,
\end{align}
and by \eqref{i_{e_0}F},
\begin{align}
\left\|i_{e_0}F\right\|_{H^{k-1}}\leq C(\left\|D\Psi\right\|_{H^{k-1}}+\left\|\Psi\right\|_{H^{k-1}}\left\|D\log N\right\|_{H^{k-1}}+\sqrt{E_k}).
\end{align}
Combining these estimates finishes the proof of the lemma.
\end{proof}
\begin{lem}\label{lem dtpsi}
As long as $\left\|D\log N\right\|_{H^k}+\left\|D\Phi\right\|_{H^{k-1}}$ is small enough
and $\left\|\Pi\right\|_{H^{k-2}}+\left\|\Phi^{-1}\right\|_{H^{k-2}}$ is uniformly bounded
, the function $\Psi$ satisfies the estimate
\begin{equation}
\begin{split}
\left\|\partial_{e_0}\Psi\right\|_{H^{k}}&\leq C(
\left\|\Pi\right\|_{H^{k-2}}+	\left\|S\right\|_{H^{k-2}}	+
\left\|D\log N\right\|_{H^{k-2}}+	\left\|D\Phi\right\|_{H^{k-2}}+	\left\|\partial_{e_0}\Phi\right\|_{H^{k-2}}\\
&\quad 	+\left\|\mathcal{L}_{e_0}\Pi\right\|_{H^{k-2}}+	\left\|\mathcal{L}_{e_0}S\right\|_{H^{k-2}}	+
\left\|D\partial_{e_0}\log N\right\|_{H^{k-2}}
)(\sqrt{E_k}+\left\|\Psi\right\|_{H^{k}}).
\end{split}
\end{equation}
\end{lem} 
\begin{proof}
By differentiating the first equation in Lemma \ref{form-eq-1} in the direction of $e_0$ and using elliptic regularity, we obtain
\begin{equation}
\begin{split}
\left\|\partial_{e_0}\Psi\right\|_{H^{k}}&\leq
C(\left\|[\mathcal{L}_{e_0},\Delta_g]\Psi\right\|_{H^{k-2}}+\left\|\partial_{e_0}(\div_g(\Psi d\log N))\right\|_{H^{k-2}}\\&\quad +
\left\|\partial_{e_0}([\mathcal{L}_{e_0},\div_g]\omega)\right\|_{H^{k-2}}
+\left\|\partial_{e_0}(g^{ij}F_{i 0}\Phi^{mp}\partial_{j}\Phi_{mp})\right\|_{H^{k-2}}).
\end{split}
\end{equation}
By using Lemma \ref{variations} and standard estimates, we get
\begin{equation}
\begin{split}
	\left\|[\mathcal{L}_{e_0},\Delta_g]\Psi\right\|_{H^{k-2}}&\leq
	C(\left\|\Pi\right\|_{H^{k-2}}(1+\left\|D\log N\right\|_{H^{k-2}})\left\|\Psi\right\|_{H^{k}}\\
	&\quad 
	+\left\|D\log N\right\|_{H^{k-2}}\left\|\partial_{e_0}\Psi\right\|_{H^{k-1}}
	+\left\|S\right\|_{H^{k-2}}\left\|\Psi\right\|_{H^{k-1}}),\\
	\left\|\partial_{e_0}(\div_g(\Psi d\log N))\right\|_{H^{k-2}}&\leq
	C[\left\|\Pi\right\|_{H^{k-2}}\left\|\Psi\right\|_{H^{k-1}}\left\|D\log N\right\|_{H^{k-1}}
	+(1+\left\|D\log N\right\|_{H^{k-2}})\cdot \\
	&\quad(\left\|\partial_{e_0}\Psi\right\|_{H^{k-1}}\left\|D\log N\right\|_{H^{k-1}}+\left\|\Psi\right\|_{H^{k-1}}\left\|D\partial_{e_0}\log N\right\|_{H^{k-1}})\\
	&\quad
	+\left\| D\log N\right\|_{H^{k-2}}(
	\left\|  S\right\|_{H^{k-2}}+\left\| \Pi\right\|_{H^{k-2}}\left\| D\log N\right\|_{H^{k-2}}
	)\left\| \Psi\right\|_{H^{k-2}}
	],\\
	\left\|\partial_{e_0}([\mathcal{L}_{e_0},\div_g]\omega)\right\|_{H^{k-2}}
	&\leq C[\left\|\Pi\right\|_{H^{k-2}}(\left\|\Pi\right\|_{H^{k-2}}+\left\|D\log N\right\|_{H^{k-2}}+\left\|S\right\|_{H^{k-2}}\\
	&\quad+\left\|\Pi\right\|_{H^{k-2}}\left\|D\log N\right\|_{H^{k-2}})\sqrt{E_k}+(\left\|\mathcal{L}_{e_0}\Pi\right\|_{H^{k-2}}\\
	&\quad+
	(1+\left\|D\log N\right\|_{H^{k-2}})\left\|\Pi\right\|_{H^{k-1}}^2+\left\|\Pi\right\|_{H^{k-2}}+\left\|D\partial_{e_0}\log N\right\|_{H^{k-2}}\\
	&\quad +\left\|\mathcal{L}_{e_0}S\right\|_{H^{k-2}}+\left\|S\right\|_{H^{k-2}}+
	\left\|\mathcal{L}_{e_0}\Pi\right\|_{H^{k-2}}\left\|D\log N\right\|_{H^{k-2}}\\
	&\quad+
	\left\|\Pi\right\|_{H^{k-2}}\left\|D\partial_{e_0}\log N\right\|_{H^{k-2}}
	+\left\|\Pi\right\|_{H^{k-2}}\left\|D\log N\right\|_{H^{k-2}}
	)\sqrt{E_k}
	\\
	&\quad+\left\|D\log N\right\|_{H^{k-2}}\left\|\mathcal{L}_{e_0}(\mathcal{L}_{e_0}\omega)\right\|_{H^{k-2}} ],
	\\
	\left\|\partial_{e_0}(g^{ij}F_{i 0}\Phi^{mp}\partial_{j}\Phi_{mp})\right\|_{H^{k-2}}
	&\leq C(
	\left\| \Pi\right\|_{H^{k-2}}\left\| i_{e_0}F\right\|_{H^{k-2}}\left\| \Phi^{-1}\right\|_{H^{k-2}}\left\| D\Phi\right\|_{H^{k-2}}\\
	&\quad +\left\| \mathcal{L}_{e_0}i_{e_0}F\right\|_{H^{k-2}}\left\| \Phi^{-1}\right\|_{H^{k-2}}\left\| D\Phi\right\|_{H^{k-2}}\\
	&\quad +\left\|i_{e_0}F\right\|_{H^{k-2}}\left\| \Phi^{-1}\right\|_{H^{k-2}}^2\left\| D\Phi\right\|_{H^{k-2}}\left\| \partial_{e_0}\Phi\right\|_{H^{k-2}}\\
	&\quad +\left\|i_{e_0}F\right\|_{H^{k-2}}\left\| \Phi^{-1}\right\|_{H^{k-2}}\left\|D \partial_{e_0}\Phi\right\|_{H^{k-2}}
	) 	.
	\end{split}
	\end{equation}
	Using the smallness of 	$\left\|D\log N\right\|_{H^k}+\left\|D\Phi\right\|_{H^{k-1}}$, we can absorb the terms containing norms of  $\partial_{e_0}\Psi$ into the left hand side of the equation. Consequently, by assuming in addition that $\left\|\Pi\right\|_{H^{k-2}}+\left\|\Phi^{-1}\right\|_{H^{k-2}}$ is uniformly bounded, we get
	\begin{equation}
	\begin{split}
	\left\|\partial_{e_0}\Psi\right\|_{H^{k}}&\leq C(\left\|\Pi\right\|_{H^{k-2}}+\left\|S\right\|_{H^{k-2}}+\left\|D\partial_{e_0}\log N\right\|_{H^{k-2}})	\left\|\Psi\right\|_{H^{k}}\\
	&\quad +C(\left\|\Pi\right\|_{H^{k-2}}+\left\|S\right\|_{H^{k-2}}+\left\|D\partial_{e_0}\log N\right\|_{H^{k-2}}+
	\left\|\mathcal{L}_{e_0}S\right\|_{H^{k-2}}+\left\|\mathcal{L}_{e_0}\Sigma\right\|_{H^{k-2}})\sqrt{E_k}	\\ &\quad                                                               +(\left\|D\log N\right\|_{H^{k-2}}+ \left\|D\Phi\right\|_{H^{k-2}}  )\left\|\mathcal{L}_{e_0}(\mathcal{L}_{e_0}\omega)\right\|_{H^{k-2}}
	\\ &\quad 
	+ (\left\|\Psi\right\|_{H^{k-1}}+\sqrt{E_k})(\left\|D\Phi\right\|_{H^{k-1}}+\left\|\partial_{e_0}\Phi\right\|_{H^{k-1}}).
	\end{split}
	\end{equation}
	Using the smallness assumptions again and treating $\left\|\mathcal{L}_{e_0}(\mathcal{L}_{e_0}\omega)\right\|_{H^{k-2}}$ by the second equation in Lemma \ref{form-eq-1} and standard estimates, we arrive at the estimate of the lemma.	
\end{proof}
\begin{prop}\label{prp-A}
	We have the energy estimate
	\begin{equation}
	\begin{split}
	\partial_TE_k&\leq C(
	\left\|\Sigma\right\|_{H^{k-1}}+	\left\|\mathrm{div}\Sigma\right\|_{H^{k-1}}	+
	\left\|N-3\right\|_{H^{k}}+	\left\|D\Phi\right\|_{H^{k-1}}\\
	&\quad 	+		\left\|\mathcal{L}_{e_0}\Sigma\right\|_{H^{k-2}}+	\left\|\mathcal{L}_{e_0}\mathrm{div}\Sigma\right\|_{H^{k-2}}	+
	\left\|\partial_{e_0} N\right\|_{H^{k}}+	\left\|\partial_{e_0}\Phi\right\|_{H^{k-1}}
	)E_k
	\end{split}
	\end{equation}
	as long as the norms of the appearing objects are uniformly bounded and $N$ is uniformly positive.
\end{prop}

\subsection{Energy estimates for the functions}
With respect to the metric $h$, given by \eqref{metric-h}, equation \eqref{wave-eq} is
\begin{align}
\Box_{h}\Phi_{mn}-2N^{-1}\partial_{e_0}\Phi_{mn}=\Phi^{pq}D_{\alpha}\Phi_{mp}D_{\beta}\Phi_{nq}h^{\alpha\beta}+\sqrt{\det\Phi}(\tau_0)^2e^{-2T} F_{m,\alpha\beta}F_{n,\gamma\delta}h^{\alpha\gamma}h^{\beta\delta}=:(*)
\end{align}
and with respect to the future-directed timelike unit normal $e_0$, we can express this equation as
\begin{align}\begin{split}\label{fct_equation}
-\partial_{e_0}(\partial_{e_0}\Phi_{mn})+\Delta_g\Phi_{mn}-\frac{2}{3}\partial_{e_0}\Phi
&=-g^{ij}\partial_i(\log N)\partial_j\Phi_{mn}-\tr_g\Pi\cdot \partial_{e_0}\Phi_{mn}-2(\frac{1}{3}-\frac{1}{N})\partial_{e_0}\Phi_{mn}+(*)\\
&=:(**)+(*),
\end{split}
\end{align}
where $(**)$ are the linear error terms and $(*)$ are the nonlinear error terms. 
The global existence of a similar system, namely wave maps from a large class of expanding spacetimes, has been
studied in \cite{BK17}.

To write down the right energy, we consider the model ODE
\begin{align}\label{model_ODE}
\ddot{X}+\frac{2}{3}\dot{X}+\lambda X=0.
\end{align}
Let
\begin{equation}
\begin{split}
\alpha=\begin{cases} 1 & \text{ if }\lambda_0>\frac{1}{9},\\
1-\sqrt{1-9\lambda_0} & \text{ if } 0<\lambda_0<\frac{1}{9}
\end{cases},
\qquad\qquad
c_E=\begin{cases} 1 & \text{ if }\lambda_0>\frac{1}{9},\\
9\lambda_0 & \text{ if } 0<\lambda_0<\frac{1}{9}
\end{cases}
\end{split}
\end{equation}
Define
\begin{align}\label{energy_ODE}
E=\frac{1}{2}(\dot{X})^2+\frac{\lambda}{2}X^2+\frac{c_E}{3}X\dot{X}.
\end{align}
\begin{lem}\label{lemma_ODE_energy}
	Let $\lambda_0$ be positive and $\lambda_0\neq \frac{1}{9}$. Then, $E$ is positive definite and if \eqref{model_ODE} holds for $\lambda\geq \lambda_0$, $\dot{E}\leq -2\alpha E$.
\end{lem}
\begin{proof}
	It is straightforward to check that $E$ is positive definite. A computation yields
	\begin{align}
	\dot{E}=(-\frac{1}{2}+\frac{c_E}{3})(\dot{X})^2-\frac{c_E}{6}X\dot{X}-\frac{c_E}{3}\lambda X^2.
	\end{align}
	In the case $\lambda_0>\frac{1}{9}$, the right hand side equals $-\frac{2}{3}\alpha E$. In the other case,
	we get $\dot{E}=-\frac{2}{3}\alpha E+Q(\dot{X},X)$, where $Q$ is a quadratic form in $(\dot{X},X)$ which is negative semidefinite. For details, see \cite[Lemma 6.4]{AnMo11} in a similar case.
\end{proof}
We denote the mean value of $\Phi_{mn}$ by $\overline{\Phi}_{mn}=\fint_M \Phi_{mn} dV$ and we write $\Phi^{\perp}=\Phi-\overline{\Phi}$. We define
\begin{align}\label{en-Phi}
E_k(\Phi)=\sum_{m,n}\sum_{l=0}^{k-1}\int_M [(-\Delta_g)^{l}\partial_{e_0}\Phi_{mn}\cdot \partial_{e_0}\Phi_{mn}+\frac{1}{2} (-\Delta_g)^{l+1}\Phi_{mn}^{\perp}\cdot \Phi_{mn}^{\perp}+\frac{c_E}{3}(-\Delta_g)^{l}\partial_{e_0}\Phi_{mn}\cdot  \Phi_{mn}^{\perp}] dV.
\end{align}
By decomposing into a basis of Laplace eigenfunctions, one sees that $E_k(\Phi)\approx \left\|\Phi^{\perp}\right\|_{H^k}+\left\|\partial_{e_0}\Phi\right\|_{H^{k-1}}$.
\begin{lem}\label{energyestimate_functions}
Suppose that \eqref{fct_equation} holds. Then, assuming that $\left\|N\right\|_{H^{k+1}}$ is uniformly bounded and $N$ is uniformly positive, we obtain the energy estimate 
\begin{equation}
\begin{split}
\partial_TE_k(\Phi)&\leq -2\alpha E_k+C(\left\|\Pi\right\|_{H^{k-1}}+\left\|S\right\|_{H^{k-1}}+\left\|N-3\right\|_{H^{k+1}})E_k+C\left\|\Phi^{-1}\right\|_{H^{k-1}}E_k^{3/2}\\
&\quad+Ce^{-2T}\left\|\sqrt{\det\Phi}\right\|_{L^{\infty}}\left\|F\right\|_{H^{k-1}}^2\sqrt{E_k}
-\frac{c_E}{3}\sum_{m,n}\partial_{e_0}\overline{\Phi}_{mn}\int_M \partial_{e_0}\Phi_{mn}dV.
\end{split}
\end{equation}
\end{lem}
\begin{proof}
We consider an arbitrary summand of the energy. For convenience, we write $u=\Phi_{mn}$.
For the rest of the proof, let $l$ be even, the odd case is similar. By integration by parts
\begin{equation}
\begin{split}
\int_M& [(-\Delta_g)^{l}\partial_{e_0}u\cdot \partial_{e_0}u+\frac{1}{2} (-\Delta_g)^{l+1}u^{\perp}\cdot u^{\perp}+\frac{c_E}{3}(-\Delta_g)^{l}\partial_{e_0}u\cdot  u^{\perp}] dV	\\
=
\int_M& [(-\Delta_g)^{l/2}\partial_{e_0}u\cdot (-\Delta_g)^{l/2}\partial_{e_0}u+\frac{1}{2} \langle D(-\Delta_g)^{l/2}u^{\perp},D(-\Delta_g)^{l/2}u^{\perp}\rangle\\
&+\frac{c_E}{3}(-\Delta_g)^{l/2}\partial_{e_0}u\cdot (-\Delta_g)^{l/2} u^{\perp}] dV=:\int_M \mathcal{E}_ldV.
\end{split}
\end{equation}
Similar as in the previous subsection, we compute
\begin{equation}
\begin{split}
\partial_T&\int_M \mathcal{E}_ldV=\int_M[ N\partial_{e_0}\mathcal{E}_l-N\mathcal{E}_l\tr_g\Pi ]dV\\
&=-\int_MN\mathcal{E}_l\tr_g\Pi dV+2\int_M N\cdot\Pi(D(-\Delta_g)^{l/2}u^{\perp}, D(-\Delta_g)^{l/2}u^{\perp})dV+ \text{Commutator terms}\\
&\quad +\int_M N(-\Delta_g)^{l/2}(\partial_{e_0}(\partial_{e_0}u))(-\Delta_g)^{l/2}\partial_{e_0}u dV + \int_M N\langle D((-\Delta)^{l/2}\partial_{e_0}(u^{\perp})),D(-\Delta)^{l/2}u^{\perp}\rangle dV\\
&\quad+\frac{c_E}{3}\int_M N((-\Delta_g)^{l/2}\partial_{e_0}u^{\perp})(-\Delta_g)^{l/2}\partial_{e_0}u + N((-\Delta_g)^{l/2}u^{\perp})(-\Delta_g)^{l/2}(\partial_{e_0}(\partial_{e_0}u)) dV.
\end{split}
\end{equation}
By integration by parts and using \eqref{fct_equation}, we can treat the last four terms as follows:
\begin{equation}
\begin{split}
&\int_M N(-\Delta_g)^{l/2}(\partial_{e_0}(\partial_{e_0}u))(-\Delta_g)^{l/2}\partial_{e_0}u dV + \int_M N\langle D((-\Delta)^{l/2}\partial_{e_0}(u^{\perp})),D(-\Delta)^{l/2}u^{\perp}\rangle dV\\
&\quad+\frac{c_E}{3}\int_M N((-\Delta_g)^{l/2}\partial_{e_0}u^{\perp})(-\Delta_g)^{l/2}\partial_{e_0}u + N((-\Delta_g)^{l/2}u^{\perp})(-\Delta_g)^{l/2}(\partial_{e_0}(\partial_{e_0}u)) dV\\
&=-\int_M N(-\Delta_g)^{l/2}((*)+(**))(-\Delta_g)^{l/2}(\partial_{e_0}u)dV
-\frac{c_E}{3}\int_M N(-\Delta_g)^{l/2}(u^{\perp})(-\Delta_g)^{l/2}((*)+(**))dV\\
&\quad- \frac{c_E}{3}\int_M N(-\Delta_g)^{l/2}(\partial_{e_0}\bar{u})((-\Delta_g)^{l/2}\partial_{e_0}u)dV
+\int_M DN*(-\Delta)^{l/2}(\partial_{e_0}u^{\perp})*D(-\Delta_g)^{l/2}u^{\perp}dV\\
&\quad +\int_M DN*(-\Delta_g)^{l/2}u^{\perp} D(-\Delta)^{l/2}u^{\perp}dV
+(-\frac{1}{2}+\frac{c_E}{3})\int_M N (-\Delta_g)^{l/2}\partial_{e_0}u\cdot (-\Delta_g)^{l/2}\partial_{e_0}u dV\\
&\quad - \frac{c_E}{3}\int_M N\langle D(-\Delta_g)^{l/2}u^{\perp},D(-\Delta)^{l/2}u^{\perp}\rangle+\frac{1}{2}
N((-\Delta_g)^{l/2}u^{\perp})((-\Delta_g)^{l/2}\partial_{e_0}u) dV\\
&\leq C(\left\|N\right\|_{L^{\infty}}\left\|(*)+(**)\right\|_{H^{k-1}}\sqrt{E_k}+\left\|DN\right\|_{L^{\infty}}E_k)\\
&\quad-2\alpha  \int_M \mathcal{E}_l dV+ C\left\|N-3\right\|_{L^{\infty}}E_k
-\frac{c_E}{3}\partial_{e_0}\bar{u}\int_M N\partial_{e_0}udV.
\end{split}
\end{equation}
In the last step we applied Lemma \ref{lemma_ODE_energy} to the last three terms before the inequality sign. Note that the last term on the right hand side only appears in the case $l=0$.
Straightforward estimates show that
\eq{\label{star-est}\alg{
\left\|(*)+(**)\right\|_{H^{k-1}}&\leq C[\left\|D\log N\right\|_{H^{k-1}}\left\|D\Phi\right\|_{H^{k-1}}+(\left\|\tr_g\Pi\right\|_{H^{k-1}}+\left\|N-3\right\|_{H^{k-1}})\left\|\partial_{e_0}\Phi\right\|_{H^{k-1}}\\
&\quad+\left\|\Phi^{-1}\right\|_{H^{k-1}}\left\|D\Phi\right\|_{H^{k-1}}^2+e^{-2T}\left\|F\right\|_{H^{k-1}}^2\left\|\sqrt{\det\Phi}\right\|_{L^{\infty}} ].
}}
It remains to consider the commutator terms. At first, we conclude from Lemma \ref{variations} by induction that for any $l\in \mathbb{N}$ and any sufficiently regular function $f$,
\begin{equation}\begin{split}
[\partial_{e_0},(\Delta_g)^l]f&=	\sum_{n=0}^{2l-2}\sum_{\sum l_i+p=n}\sum_{p=0}^n\underbrace{D^{l_1+1}\log N*\ldots D^{l_p+1}\log N}_{p-\text{times}}*\\
&\quad \sum_{m=0}^{2(l-1)-n}[D^m\Pi*D^{2l-n-m}f	+D^{2l-n-m-1}\log N*D^{m+1}\partial_{e_0}f\\
&\quad+ D^{m} S*D^{2l-n-m}f
+\sum_{m=0}^{l-n} D^{m}(\Pi*D\log N)*D^{2l-n-m-1}f
].
\end{split}
\end{equation}
Assuming that $\left\|D\log N\right\|_{H^{k}}$ is uniformly bounded, the four commutator terms can be estimated by
\begin{equation}
\begin{split}
\int_M N[\partial_{e_0},(-\Delta)^{l/2}]\partial_{e_0}u\cdot (-\Delta)^{l/2}\partial_{e_0}udV
&\leq C\left\|N\right\|_{L^\infty}[(\left\|\Pi\right\|_{H^{k-2}}+\left\|S\right\|_{H^{k-2}})\left\|\partial_{e_0}u\right\|_{H^{k-1}}\\
 &\quad+\left\|D\log N\right\|_{H^{k-1}}\left\|\partial_{e_0}(\partial_{e_0}u)\right\|_{H^{k-2}}]
 \left\|\partial_{e_0}u\right\|_{H^{k-1}},\\
 \int_M N\langle[\mathcal{L}_{e_0},D(-\Delta)^{l/2}]u^{\perp},D (-\Delta)^{l/2}u^{\perp}\rangle dV
 &\leq C\left\|N\right\|_{L^\infty}[(\left\|\Pi\right\|_{H^{k-1}}+\left\|S\right\|_{H^{k-1}})\left\|u^{\perp}\right\|_{H^{k}}\\
 &\quad+\left\|D\log N\right\|_{H^{k}}\left\|\partial_{e_0}u\right\|_{H^{k-1}}]
 \left\|u^{\perp}\right\|_{H^{k}},\\
 \int_M N[\partial_{e_0},(-\Delta)^{l/2}]u^{\perp}\cdot (-\Delta)^{l/2}\partial_{e_0}udV
 &\leq C\left\|N\right\|_{L^\infty}[(\left\|\Pi\right\|_{H^{k-2}}+\left\|S\right\|_{H^{k-2}})\left\|u^{\perp}\right\|_{H^{k-1}}\\
 &\quad+\left\|D\log N\right\|_{H^{k-1}}\left\|\partial_{e_0}u\right\|_{H^{k-2}}]
 \left\|\partial_{e_0}u\right\|_{H^{k-1}},\\
  \int_M N (-\Delta)^{l/2}u^{\perp}\cdot[\partial_{e_0},(-\Delta)^{l/2}]\partial_{e_0}udV
 &\leq C\left\|N\right\|_{L^\infty}[(\left\|\Pi\right\|_{H^{k-2}}+\left\|S\right\|_{H^{k-2}})\left\|\partial_{e_0}u\right\|_{H^{k-1}}\\
 &\quad+\left\|D\log N\right\|_{H^{k-1}}\left\|\partial_{e_0}(\partial_{e_0}u)\right\|_{H^{k-2}}]
 \left\|u^{\perp}\right\|_{H^{k-1}}.
\end{split}
\end{equation}
The statement now follows from combining all the estimates and using \eqref{fct_equation}.
\end{proof}
\begin{lem}\label{variation_meanvalue}
Let $u$ be a function, $\bar{u}=\fint_M udV$ and $u^{\perp}=u-\bar{u}$. Then we have
\begin{align}
\partial_{e_0}\bar{u}=N^{-1}\left(\fint_M N\partial_{e_0}udV-\fint_M \mathrm{tr}_g\Pi\cdot u^{\perp}dV\right).
\end{align}
In particular, under  the assumptions of Proposition \ref{energyestimate_functions},
\begin{align}
|\partial_{e_0}\bar{u}-\overline{\partial_{e_0}u}|\leq C\left\|N-3\right\|_{L^{\infty}}\left(\left\|\partial_{e_0}u\right\|_{L^2}+\left\|u^{\perp}\right\|_{L^2}\right).
\end{align}
\end{lem}
\begin{proof}
Recall that $\partial_tg=-2\Pi-\mathcal{L}_Xg$ and $e_0=N^{-1}(\partial_T+X)$. Then a straightforward computation shows
\begin{equation}
\begin{split}
\partial_{e_0}\bar{u}&=N^{-1}\partial_T\frac{\int_MudV}{\int_M dV}=N^{-1}\left(\frac{\int_M \partial_TudV+\frac{1}{2}\int_Mu\cdot\mathrm{tr}\partial_TgdV}{\int_MdV}-\frac{1}{2}\frac{\int_MudV\cdot\int_M\mathrm{tr}\partial_TgdV}{(\int_MdV)^2}\right)\\
&=\frac{1}{N\int_MdV}\left(\int_M\partial_{T+X}udV-\int_Mu\cdot\mathrm{tr}_g\Pi+\fint_MudV\cdot\int_M\mathrm{tr}_g\Pi dV\right).
\end{split}
\end{equation}
The second assertion of the lemma follows from standard estimates and using $\Pi=-\Sigma+(N^{-1}-3^{-1})g$.
\end{proof}
\begin{prop}\label{prp-phi}
Under the assumptions of Lemma \ref{energyestimate_functions}, we have
\begin{align}\begin{split}
\partial_TE_k(\Phi)&\leq -2\alpha E_k+C(\left\|\Sigma\right\|_{H^{k-1}}+\left\|\mathrm{div}\Sigma\right\|_{H^{k-1}}+\left\|N-3\right\|_{H^{k+1}})E_k+C\left\|\Phi^{-1}\right\|_{H^{k-1}}E_k^{3/2}\\
&\quad+Ce^{-2T}\left\|\sqrt{\det\Phi}\right\|_{L^{\infty}}\left\|F\right\|_{H^{k-1}}^2\sqrt{E_k}.
\end{split}
\end{align}	
\end{prop}
\begin{proof}
This follows from Proposition \ref{energyestimate_functions}, Lemma \ref{variation_meanvalue},
and using the notations $S=\mathrm{div}_g\Pi-D\mathrm{tr}_g\Pi$ and $\Pi=-\Sigma+(N^{-1}-3^{-1})g$.
\end{proof}

\begin{rem}
Note that we have control over \(\left\|\Phi^{-1}\right\|_{H^{k-1}}\) due to Lemma \ref{variation_meanvalue} 
and the exponential decay of the energy that we will obtain.
\end{rem}
\section{Elliptic estimates} \label{sec : ell}
We provide in this section the standard elliptic estimates for lapse and shift and their time-derivatives.
\begin{prop}\label{prop-ell-est}
Under smallness conditions for the lapse function, a pointwise estimate of the form $0<N\leq 3$ holds and moreover the following two estimates.
\eq{\alg{
\Abk{N-3}\ell&\leq C\left(\Abk{\Si}{\ell-2}^2+\ab\tau\Abk{\rho}{\ell-2}+\tau^3 \Abk{\underline{\eta}}{\ell-2}\right),\\
\Abk{X}\ell&\leq C\left(\Abk{\Si}{\ell-2}^2+\Abk{g-\gamma}{\ell-1}^2+\ab\tau\Abk{\rho}{\ell-3}+\tau^3 \Abk{\underline{\eta}}{\ell-3}+\tau^2 \Abk{N\jmath}{\ell-2}\right)
}}
\end{prop}
\begin{proof}
These estimates are an immediate consequence of elliptic regularity applied to \eqref{lapse} and \eqref{shift}, respectively and the maximum principle applied to \eqref{lapse}.
\end{proof}
\subsection{Estimates of the time derivatives}
\begin{lem}
Let $\ell\geq 4$. For sufficiently small perturbations, the following estimate holds.
\eq{\alg{
\Abk{\p TN}\ell&\leq C\Big[\Abk{\widehat N}\ell+\Abk{X}{\ell+1}+\Abk{\Si}{\ell-1}^2+\Abk{g-\gamma}\ell^2+\ab{\tau}\Abk{S}{\ell-2}\\
&\qquad\quad+\ab{\tau}\Abk{\rho}{\ell-1}+\ab\tau^3\Abk{\underline\eta}{\ell-2} +\ab\tau^2\Abk{\jmath}{\ell-1}+\ab\tau^3\Abk{\underline T}{\ell-1}\\
&\qquad\quad+\tau^2\Abbk{F}{\ell-2}\Big(\Abk{\mcr L_{e_0}\omega}{\ell-2}+\Abk{\omega}{\ell-1}+\Abbk{F}{\ell-2}\Big)+\left(\Abk{\p T\Phi}{\ell-2}+\Abk{D\Phi}{\ell-2}\right)^2\Big],\\
\Abk{\p TX}\ell&\leq C\Big[\Abk{X}{\ell+1}+\Abk{\Si}{\ell-1}^2+\Abk{g-\gamma}{\ell}^2+\Abk{\widehat N}{\ell}+\ab\tau\Abk{S}{\ell-2}+\ab{\tau}\Abk{\rho}{\ell-1}+\ab\tau^3\Abk{\underline\eta}{\ell-2}\\
&\qquad\quad+\ab\tau^2\Abk{\jmath}{\ell-1}+\ab\tau^3\Abk{\underline T}{\ell-1}+\tau^2\Abbk{F}{\ell-2}\Big(\Abk{\mcr L_{e_0}\omega}{\ell-2}+\Abk{\omega}{\ell-1}+\Abbk{F}{\ell-2}\Big)\\
&\qquad+\left(\Abk{\p T\Phi}{\ell-2}+\Abk{D\Phi}{\ell-2}\right)^2\Big]
}}
The constant $C$ depends implicitly on the perturbation via 
\eq{
C=C(\Abk{X}{\ell+1},\Abk{N-3}{\ell+1},\Abi{N},\Abi{N^{-1}},\Abk{\Si}{\ell-1},\Abk{g-\gamma}{\ell}).
}
\end{lem}

\begin{proof}
By differentiation with respect to $T$ the elliptic system implies 
\eq{\label{tdlapse}\alg{
\/\left(\Delta-\frac13\right)\p TN&= 2N\langle D D N,\Si\rangle-2\widehat N\Delta N+\langle D D N,\mcl L_Xg\rangle\\
&\quad+\left(2 D^k(N\Si_k^i)+ D^i(\widehat N)-\frac12\Delta X^i-\frac12 D^k D^iX_k\right) D_i N\\
&\quad+2N\Big(-2N\abg{\Si}^3+2\widehat N\abg{\Si}^2-2\langle D X,\Si,\Si\rangle-2\abg{\Si}^2\\
&\qquad\qquad\,\,-N\langle\Si,\frac12\mcl L_{g,\ga}(g-\gamma)+J\rangle+\langle\Si, D D N\rangle+2N\abg\Si^3-\widehat N\abg\Si^2\\
&\qquad\qquad\,\,-2\langle\Si,\mcl L_Xg\rangle+8\pi\ab\tau\langle\Si,S\rangle\Big)\\
&\quad+N\Big(\p T(\ab{\tau}\rho)+\p T(\ab{\tau}^3\underline\eta)\Big)+\left(\abg{\Si}^2+\ab{\tau}\rho+\ab{\tau}^3\underline{\eta}\right)\p TN,
}}
where $\langle{DX,\Si,\Si}\rangle = D_iX^j\Si^i_k\Si^k_j$ and
\eq{\label{tdshift}\alg{
\Delta (\p T X^i)+R_m^i(\p T X^m)&=-(\p TR^i_m)X^m-[\p T,\Delta]X^i\\
&\quad+2 D_j(\p T N)\Si^{ij}+2 D_j N(\p T\Si^{ij})-(\p Tg^{ik}) D_k\widehat N-\frac13g^{ik}D_k(\p TN)\\
&\quad+2(\p TN)\ab\tau^{2}\jmath^b+2N\p T(\ab\tau^{2}\jmath^b)\\
&\quad- 2(\p TN)\Si^{mn}(\Gamma_{mn}^i-\widehat{\Gamma}_{mn}^i)- 2N(\p T\Si^{mn})(\Gamma_{mn}^i-\widehat{\Gamma}_{mn}^i)\\
&\quad- 2N\Si^{mn}\p T\Gamma_{mn}^i+ (\p Tg^{mk}g^{nl}) D_kX_l(\Gamma_{mn}^i-\widehat{\Gamma}_{mn}^i)\\
&\quad+  D^m(\p TX^n)(\Gamma_{mn}^i-\widehat{\Gamma}_{mn}^i)
+  D^mX^n\p T\Gamma_{mn}^i.
}}
We proceed analogous to \cite{AnFa17} using the evolution equations for the energy-density and the current, which are independent of the matter model. 
The divergence identity of the energy momentum tensor in the unrescaled form, $\widetilde{\nabla}_{\al}\widetilde T^{\al\be}$ (cf.~\cite{Re08}, (2.66), (2.67)) reads with respect to the rescaled variables, $\rho=\tilde{\rho}\ab\tau^{-3}$ and $\jmath=\ab\tau^{-5}\tilde \jmath$, 
\eq{\label{tdrhoeta}\alg{
\p T\rho&=(3-N) \rho-X^i\na_i\rho+\tau N^{-1}\na_i(N^2\jmath^i)-\tau^2\frac N3g_{ij}T^{ij}-\tau^2N\Si_{ij}T^{ij},\\
\p T\jmath^i&=\frac53(3-N)\jmath^i-X^j\na_j\jmath^i-(\na^iX_j)\jmath^j+\tau\na_j(N T^{ij})-2N\Si_j^i\jmath^j-\ab\tau^{-1}\rho \na^iN.
}}
 The time derivative of the term containing $\eta$, however, requires a detailed evaluation as it depends on the equations of motion for the matter model. We need to estimate the $H^{\ell-2}$-norm of $\p T (\tau^3\underline \eta)$. This term is $\tau^3\underline \eta=4\pi\tau^{-2}\tilde g^{ij}\tilde T_{ij}$ and up to a constant and the factor $\tau^{-2}$ it is evaluated in \eqref{s-trace}. We now take the time derivative of the terms on the right-hand side of \eqref{s-trace} modulo the $\tau^2$-factor and replace, if necessary, second time derivatives of the matter fields using the corresponding equations of motion. We do this explicitly for two terms to illustrate the computation and leave the remaining terms to the reader. This computation will provide an estimate for $\Abk{\p T\tau^3\underline \eta}{\ell-2}$. The first term we consider explicitly is
\eq{\alg{
\p T\left[g^{ij}\left(\sqrt{\det \Phi}\tr\left(F^\mu_jF_{i\mu}\right)\right)\right]&=
(\p Tg^{ij})\left(\sqrt{\det \Phi}\tr\left(F^\mu_jF_{i\mu}\right)\right)+(\p T\sqrt{\det \Phi}) g^{ij} \tr\left(F^\mu_jF_{i\mu}\right)\\
&\quad+g^{ij}\sqrt{\det \Phi}\left(\p T\tr\left(F^\mu_jF_{i\mu}\right)\right).
}}
The norms of the first terms on the right-hand side can directly be estimated. We focus on the evaluation of the last term.
\eq{\alg{
g^{ij}\p T \tr (F^\mu_jF_{i\mu})&=g^{ij}\delta^{mn}\p T \Big(\tau^2\Big[N^{-2}(\tau F_{j0,m})(\tau F_{i0,n})+N^{-2}X^k\left(\tau F_{jk,m}F_{i0,n}+\tau F_{j0,m}F_{ik,n}\right)\\
&\quad+(g^{kl}-\hat X^{k}\hat X^l)F_{jk,m}F_{il,n}\Big]\Big)=g^{ij}\delta^{mn} \tau^2 N^{-2}(\tau F_{j0,m})\p T(\tau F_{i0,n})+\hdots
}} 
Here we suppress terms that can either directly be estimated or those that can be handled similarly to the one considered explicitly. We proceed with that term.
\eq{\alg{
g^{ij}\delta^{mn} \tau^2 N^{-2}(\tau F_{j0,m})\p T(\tau F_{i0,n})=-g^{ij}\delta^{mn} \tau^2 N^{-2}(\tau F_{j0,m})\left(\p i\p T A_{T,n}-\p T^2A_{i,n}\right)
}}
There are two terms with time derivatives on the right-hand side, which cannot be estimated by the energies. We therefore replace those by the corresponding evolution equations or by suitable quantities estimated in the respective sections on the control of the matter fields. The relation between $A$, $\Psi$ and $\omega$ and the definition of $e_0$ imply
\eq{\alg{
\p T A_{T,n}&= N\p{e_0}\Psi_n-X\Psi_n+(\p TN)\Psi_n - (\p TA_{i,n}) X^i-A_{i,n } \p TX^i,\\
\p T^2 A_{\ell,n}&= \mcl L_{e_0}\mcl L_{e_0}\omega_{\ell,n}+N^{-3}(\p TN+X^i\p i N)\cdot (\p T+X^i\p i)\omega_{\ell,n}-(X^i\p i)^2 \omega_{\ell,n}-X^i\p i\p T\omega_{\ell,n}\\
&\quad -(\p TX^i)\p i\omega_{\ell,n}-X^i\p T\p i\omega_{\ell,n}-\p k(\hat X^j)N^{-1}(\p T+X^i\p i)\omega_{j,n}-\mcl L_{e_0}(\p \ell \hat X^i)\omega_{i,n}.
}}
We intend to use those equations to replace the left-hand side appearing in the time-differentiated equations by the right-hand sides, which can then be estimated using the corresponding results from section \ref{sec : 62}. We can estimate the Sobolev norms of $\Psi$ and $\p T\Psi$ by Lemmas \ref{lem psi} and \ref{lem dtpsi}, respectively. The spatial components, i.e.~$\omega$ and their time-derivatives are estimated using equivalency of energies as stated in \eqref{energy-vector-potential}. The term $\mcl L_{e_0}\mcl L_{e_0}\omega_{\ell,n}$ is substituted using equation \eqref{form-eq-12}. This, in turn, makes terms in $\Phi$ appear, for which we use the standard Sobolev norm. Proceeding as described leads to an estimate for the $H^{\ell-2}$-norm of the left-hand side by
\eq{\alg{
&C \tau^2 \Abbk F {\ell-2} \cdot \Bigg\{\Abk{N\p{e_0}\Psi_n-X\Psi_n+(\p TN)\Psi_n - (\p TA_{i,n}) X^i-A_{i,n } \p TX^i}{\ell-2}\\
&\quad+ \Abk{\mcl L_{e_0}\mcl L_{e_0}\omega_{\ell,n}+N^{-3}(\p TN+X^i\p i N)\cdot (\p T+X^i\p i)\omega_{\ell,n}-(X^i\p i)^2 \omega_{\ell,n}-X^i\p i\p T\omega_{\ell,n}}{\ell-2}\\
&\quad+\Abk{-(\p TX^i)\p i\omega_{\ell,n}-X^i\p T\p i\omega_{\ell,n}-\p k(\hat X^j)N^{-1}(\p T+X^i\p i)\omega_{j,n}-\mcl L_{e_0}(\p \ell \hat X^i)\omega_{i,n}}{\ell-2}\Bigg\}\\
&\qquad\quad\leq C \tau^2 \Abbk F {\ell-2} \cdot \Bigg\{ (1+\Abk{\p TN}{\ell-2}+\Abk{\p TX}{\ell-1}) \left(\Abk{\mcr L_{e_0} \omega}{\ell-2}+\Abk{\omega}{\ell-1}\right)\\
&\qquad\qquad\qquad\qquad\qquad\qquad+\Abbk{F}{\ell-2} \left(\Abk{N-3}{\ell-1}+\Abk{D\Phi}{\ell-2}+\Abk{\Pi}{\ell-2}+\Abk{\p T\Phi}{\ell-2}\right)\\
&+\left(\Abk{S}{\ell-3}+\Abk{D\Phi}{\ell-3}+\Abk{\p T\Phi}{\ell-3}+\Abk{\mcr L_{e_0}\Si}{\ell-3}+\Abk{\mcr L_{e_0}\div \Si}{\ell-3}\right)\left(\Abk{\mcr L_{e_0} \omega}{\ell-2}+\Abk{\omega}{\ell-1}\right)\Bigg\},
}}
where $C=C(\Abi N,\Abi {N^{-1}},\Abi{\Phi},\Abk{X}{\ell-1},\Abk{N-3}{\ell-1},\Abk{\Pi}{\ell-2},\Abk{D\Phi}{\ell-3})$. The second term from \eqref{s-trace} that we estimate explicitly is
\eq{\alg{
&\Abk{\p T\left(\frac{1}{N^2}\p T\Phi^{mq}\p T\Phi_{mq}\right)}{\ell-2}\leq C\Big\{ \Abk{\p T N}{\ell-2}\Abk{\p T\Phi}{\ell-2}^2\\
&+\Abk{\p T\Phi}{\ell-2}\Bigg[ (\Abk{\p T\Phi}{\ell-2}+\Abk{D\Phi}{\ell-2}) (\Abk{\p TN}{\ell-2}+\Abk{N-3}{\ell-1}+\Abk{\p TX}{\ell-2}+\Abk{X}{\ell-1})\\
&\qquad\qquad\qquad+\Abk{D\Phi}{\ell-1}+\Abk{\partial_{e_0}\Phi}{\ell-2}+\Abk{(**)}{\ell-2}+\Abk{(*)}{\ell-2}\Bigg]\Big\}.
}}
Note that the last two terms are defined in \eqref{fct_equation} and estimated in \eqref{star-est}. Evaluating the remaining terms of \eqref{s-trace} after taking the time derivative we conclude an estimate of the following form.
\eq{\label{extr-pr}\alg{
&\Abk{\p T\tau^3\underline \eta}{\ell-2}\\
&\leq C\bigg\{ \Abk{\p TN}{\ell-2}\left(\tau^2\Abbk{F}{\ell-2}\left(\Abk{\mcr L_{e_0} \omega}{\ell-2}+\Abk{\omega}{\ell-1}\right)+\Abk{\p T\Phi}{\ell-2}\left(\Abk{\p T\Phi}{\ell-2}+\Abk{D\Phi}{\ell-2}\right)\right)\\
&\qquad+\Abk{\p TX}{\ell-2}\left(\tau^2\Abbk{F}{\ell-2}\left(\Abk{\mcr L_{e_0} \omega}{\ell-2}+\Abk{\omega}{\ell-1}\right)+\Abk{\p T\Phi}{\ell-2}\left(\Abk{\p T\Phi}{\ell-2}+\Abk{D\Phi}{\ell-2}\right)\right)\bigg\}\\
&\qquad+\hdots}}
Here, the suppressed terms are not in factors of the norms of $\p TN$ and $\p TX$ and therefore contribute directly to the right-hand side of the final elliptic estimate for the norm of $\p TN$. The terms, which are listed explicitly are handled in the following way. We note that the factor multiplied with the term $\Abk{\p TN}{\ell-2}$ is small by assumption. Applying elliptic regularity to \eqref{tdlapse} this term can therefore be absorbed in the constant. \\
On the right-hand side of \eqref{extr-pr} a term with $\Abk{\p TX}{\ell-2}$ remains, which is also multiplied by a small factor. This preliminary estimate for $\p TN$ is then used in conjunction with the elliptic estimate for \eqref{tdshift}, which contains $\p T N$ terms that are replaced by the preliminary estimate. This estimate in turn contains $\p T X$ terms on the right-hand side, which can be absorbed using smallness of the factors and we obtain an estimate for $\Abk{\p TX}{\ell}$ independent of $\p TN$. This can then in turn be used in the preliminary estimate for $\Abk{\p TN}{\ell}$ to obtain the final estimate for $\Abk{\p TN}{\ell}$.
\end{proof}


\section{Proof of the main theorem} \label{sec : pro}

\subsection{Preliminaries and local existence}
Small perturbations of an initial data set, corresponding to the background solution, are not necessarily CMC. As argued in \cite{FK15}, for a related situation, the corresponding maximal globally hyperbolic development contains a CMC surface with data close to the background. A similar argument applies in the present context. Starting from this CMC surface we apply the local existence theory for the reduced system, which is of hyperbolic-elliptic nature. An analysis as in \cite{AnMo03} yields a local existence theory for our system and a continuation criterion assuring the existence as long as the Sobolev norms in suitable regularity ($H^4$ for metric and fields and $H^3$ for time derivatives) is sufficient. It therefore suffices to establish the energy decay to conclude global existence. We now give a detailed description of the main theorem.
\begin{thm}\label{thm-1-version2}
	Let $(M,\gamma)$ be a compact, negative, 3-dimensional Einstein manifold without boundary and Einstein constant $\mu=-\frac29$ and $\Phi_{\mathsf b}$ a set of constant functions corresponding to a flat metric on $\mathbb T^q$ . Then there exists an $\varepsilon>0$ such that for an rescaled initial data set $(g,\Si,A,\dot A,\Phi,\dot \Phi)\in H^4\times H^3\times H^4\times H^3\times H^4\times H^3$ with
	\eq{
		(g,\Si,A,\dot A,\Phi,\dot\Phi)\in \mathbf B_{\varepsilon} \left(\gamma,0,0,0,\Phi_{\mathsf b},0\right)
	}
	the corresponding solution to the rescaled Einstein-Kaluza-Klein system \eqref{maxwell-eq} - \eqref{conformal system},  is future-global in time and future complete. As the mean curvature $\tau$ of the macroscopic part tends to zero the perturbation $(g-\gamma,\Si,A,\Phi-\Phi_{\infty})$ goes to zero in $H^4\times H^3\times H^4\times H^4$. In particular, the field $\Phi$ asymptotically freezes, i.e.
	\eq{
		(g,\Si,A,\Phi)\rightarrow (\gamma,0,0,\Phi_{\infty})
	} 
	for some set $\Phi_{\infty}$ of constant functions. In particular, the Milne model is an attractor for the macroscopic geometry of product spacetimes with a torus as an internal space within the class of perturbations that preserve the full symmetry group of the torus.
\end{thm}
\subsection{Global existence}
The proof of Theorem \ref{thm-1-version2} is an almost immediate consequence of the individual energy estimates for the geometry, the one-forms and the functions. We define a total energy measuring all perturbations simultaneously by
\eq{
\mathbf E_{\mathsf{tot}}(g-\gamma,\Si,\omega,\Phi):= E_4(g-\gamma,\Si)+e^{-2T}E_4(\omega)+E_4(\Phi).
}

The energy-estimate for the total energy is given by

\begin{lem}
Under the smallness assumption on the perturbation the following estimate holds.
\eq{
\label{enery-ode-inequality}
\p T\mathbf E_{\mathsf{tot}}(g-\gamma,\Si,\omega,\Phi)\leq -2\alpha \mathbf E_{\mathsf{tot}}(g-\gamma,\Si,\omega,\Phi)+C\mathbf E_{\mathsf{tot}}(g-\gamma,\Si,\omega,\Phi)^{3/2}
}
\end{lem}

\begin{proof}
This estimate is a consequence of Propositions \ref{en-est-geom-impr}, \ref{prp-A} and \ref{prp-phi} and the elliptic estimates.
\end{proof} 
In order to determine the decay rate of the total energy let us consider the model
equation \[\partial_T y(T)=-2\alpha y(T)+Cy(T)^\frac{3}{2}.\] 
For \(y_0:=y(0)>0\) we obtain the solution
\begin{align*}
y(T)=\frac{4\alpha^2}{\big(e^{\alpha T}(\frac{2\alpha}{\sqrt{y_0}}-C)+C\big)^2}. 
\end{align*}
If we assume that \(y_0<\frac{4\alpha^2}{C^2}\) then we can deduce that the 
solution \(y(T)\) has a decay rate of $e^{-2\alpha T}$.

Performing a similar analysis of \eqref{enery-ode-inequality} we can conclude
that the total energy decays with a rate of $e^{-2\alpha T}$. In the following, we will use this result to determine the decay of the individual energies based on the individual energy estimates. 
\subsection{Decay rates}
Appealing to the individual energy estimate for $E_4(\omega)$ this yields the following decay rates.
\begin{lem} For sufficiently small initial perturbations, the following estimates hold.
\eq{\alg{
E_4(g-\gamma,\Si)\lesssim e^{-2\alpha T},\quad E_4(\omega)\lesssim 1,\quad E_4(\Phi)\lesssim e^{-2\alpha T}
}}
\end{lem}
\subsection{Completeness}
The energy decay rates of the macroscopic geometry are identical to those of Andersson-Moncrief for the vacuum Einstein flow \cite{AnMo11}. Therefore future completeness follows analogously. This completes the proof of Theorem \ref{thm-1-version2}.
\section{Other related systems} \label{sec : rel}
In this section we list other well-known models for which our method applies. More precisely, one also obtains nonlinear stability of the $3+1$-dimensional Milne model as a solution of the following systems.
 \subsection{The Brans-Dicke system}
 The Brans-Dicke model is governed by the action
 \begin{align}
 S(h,\phi)=\int_M(\phi R-\frac{\omega}{\phi}h^{\alpha\beta}\nabla_\alpha\phi\nabla_\beta\phi)dV_h,
 \end{align}
where \(\omega\) denotes the dimensionless coupling constant. 
The critical points of the Brans-Dicke action are given by
\begin{equation}
\begin{split}
R_{\alpha\beta}-\frac{1}{2}Rh_{\alpha\beta}=&\frac{\omega}{\phi^2}(\nabla_\alpha\phi\nabla_\beta\phi-\frac{1}{2}h_{\alpha\beta}h^{\delta\gamma}\nabla_\delta\phi\nabla_\gamma\phi)
+\frac{1}{\phi}(\nabla_\alpha\nabla_\beta\phi-\Box\phi),\\
\Box\phi=&0.
\end{split}
\end{equation}
This system is obtained from our result by setting \(F=0\) and assuming $\mathbb{T}^q=S^1$ so that \(\Phi_{mn}=\phi\).

\subsection{The Einstein-wave map system}
In order to define the Einstein-wave map system we also take into account a Riemannian manifold 
\((P,k_{ij})\) and consider a map \(\phi\colon M\to P\). This allows us to provide the action
for Einstein-wave maps
\begin{align}
S(h,\phi)=\int_M(R-h^{\alpha\beta}\nabla_\alpha\phi^i\nabla_\beta\phi^jk_{ij}(\phi))dV_h.
\end{align}
The critical points of the Einstein-wave map system are given by
\begin{equation}
\begin{split}
R_{\mu\nu}-\frac{1}{2}Rh_{\mu\nu}&=\nabla_\mu\phi^i\nabla_\nu\phi^jk_{ij}-\frac{1}{2}\nabla_\alpha\phi^i\nabla_\beta\phi^jh^{\alpha\beta}k_{ij}h_{\mu\nu},\\
\Box_g\phi^i&=-\Gamma^i_{jk}(\phi)\nabla_\alpha\phi^j\nabla_\beta\phi^kh^{\alpha\beta},
\end{split}
\end{equation}
where \(\Gamma^i_{jk}(\phi)\) are the Christoffel symbols on the Riemannian manifold \(P\). This system is a slight modification of our system in the case of \(F=0\) since we are now assuming that the map \(\phi\)
takes its values in a Riemannian manifold. If the map is almost constant one can assume that its
image is contained in a single coordinate chart such that we can think of it as a set of functions
rather than a map between manifolds. The Einstein-wave map system is not directly captured
by our main result by setting \(F=0\) but exactly the same energies can be used.	

\subsection{The Einstein-Maxwell system} 
To define the energy for the Einstein-Maxwell system we consider the vector potential \(A_\mu\) and 
its curvature two-form \(F_{\mu\nu}\). The energy functional for the Einstein-Maxwell system is the following
\begin{align}
S(h,F)=\int_M(R-\frac{1}{2}F_{\alpha\beta}F_{\gamma\delta}h^{\alpha\gamma}h^{\beta\delta})dV_h.
\end{align}
The critical points of the Einstein-Maxwell system are given by
\begin{align}
\begin{split}
R_{\mu\nu}-\frac{1}{2}Rh_{\mu\nu}=&F_{\mu\beta}F_{\nu\alpha}h^{\alpha\beta}-\frac{1}{4}F_{\alpha\beta}F_{\gamma\delta}h^{\alpha\gamma}h^{\beta\delta}h_{\mu\nu},\\
h^{\alpha\beta}\nabla_\alpha F_{\beta\delta}=&0.
\end{split}
\end{align}
We obtain the Einstein-Maxwell system by setting \(\Phi_{pq}=0\), neglecting the equation for \(\Phi_{pq}\) and changing some constants on the right hand side of the equation on the metric.


\begin{thebibliography}{DWW07}

\providecommand{\url}[1]{\texttt{#1}}
\expandafter\ifx\csname urlstyle\endcsname\relax
  \providecommand{\doi}[1]{doi: #1}\else
  \providecommand{\doi}{doi: \begingroup \urlstyle{rm}\Url}\fi
\bibitem[AMa]{AnMo03}
\textsc{Andersson}, L.~; \textsc{Moncrief}, V.~:
\newblock {Elliptic--hyperbolic systems and the Einstein equations }
\newblock \emph{Ann.~Henri Poincar\'e} \textbf 4, 2003 

\bibitem[AMb]{AnMo11}
\textsc{Andersson}, L.~; \textsc{Moncrief}, V.~:
\newblock {Einstein spaces as attractors for the Einstein flow.}
\newblock {In: }\emph{J. Differ. Geom.} \textbf{89} (2011),  1--47


\bibitem[AF17]{AnFa17}
 \textsc {Andersson}, L.~; \textsc{Fajman}, D.~:
\newblock{Nonlinear Stability of the Milne model with matter},
\emph{arXiv:1709.00267},
{2017}
 
\bibitem[BK]{BK17}
\textsc{Branding}, V. ; \textsc{Kr\"oncke}, K.:
\newblock \emph{Global existence of wave maps and some generalizations on expanding spacetimes},
\newblock Calc. Var. (2018) 57: 119

\bibitem[BZ]{BZ}
\textsc{Bieri}, L.; \textsc{Zipser}, N.
\newblock \emph{Extensions of the stability theorem of the Minkowski space in general relativity}
\newblock American Mathematical Society, International Press, 2009

\bibitem[CK]{CK92}
\textsc{Christodoulou}, D. ; \textsc{Klainerman}, S.:
\newblock \emph{The global nonlinear stability of the Minskowski space}



\bibitem[CM]{CM01}
\textsc{Choquet-Bruhat}, Y. ; \textsc{Moncrief}, V.:
\newblock \emph{Future Global in Time Einsteinian Spacetimes with $U(1)$ Isometry Group}
\newblock Ann.~Henri Poincar\'e, \textbf{2}, 2001

\bibitem[CH09]{CH09}
\textsc{Choquet-Bruhat}, Y. :
\newblock \emph{General Relativity and the Einstein equations}
\newblock Oxford Mathematical Monographs
\newblock Oxford Science Publications







\bibitem[FK15]{FK15}
\textsc{Fajman}, D. ; \textsc{Kr\"oncke}, K.
\newblock{Stable fixed points of the Einstein flow with a positive cosmological constant}
\newblock to appear in \emph{Commun.~Geom.~Anal.}

\bibitem[VK]{Kr16}
\textsc{Valiente Kroon}, J.~A.~:
\newblock \emph{Conformal Methods in General Relativity},
\newblock Cambridge UK,Cambridge University Press, 2016

\bibitem[Kr15]{Kr15}
\textsc{Kr\"oncke}, K.:
\newblock {On the stability of Einstein manifolds}
\newblock \emph{Ann.~Glob.~Anal.~Geom.}
\textbf{47} (2015), 81--98

\bibitem[LR]{LR}
\textsc{Lindblad}, H. ; \textsc{Rodnianski}, I.
\newblock \emph{The global stability of Minkowski space-time in harmonic gauge}
\newblock Ann.~of Math. (2), 171, 2010





\bibitem[OW]{Ow97}
\textsc{Overduin}, J. M. ; \textsc{Wesson}, P. S.:
\newblock \emph{Kaluza-{K}lein gravity},
\newblock  Phys. Rep. 283, 5-6, 1997, 303-378

\bibitem[Po05]{Po05}
\textsc{Polchinski}, J.~:
\newblock \emph{String Theory},
\newblock Volume I, An Introduction to the bosonic string,
\newblock Cambridge University Press


\bibitem[Re]{Re08}
\textsc{Rendall}, A.~D.:
\newblock \emph{{Partial Differential Equations in General Relativity}}
\newblock {Oxford Graduate Texts in Mathematics}, 2008 

\bibitem[Ri]{Ri08}{
   \textsc{Ringstr{\"o}m, Hans},
  \newblock{Future stability of the Einstein-non-linear scalar field system},
   \newblock {In: }\emph{Invent. Math.},
   \textbf{173},
   (2008), {123--208}}

\bibitem[RS18a]{RoSp18}
\textsc{Rodnianski}, I., \textsc{Speck}, J.:
\newblock \emph{A regime of linear stability for the Einsetin-scalar field system with applications to nonlinear Big Bang formation}
\newblock Ann.~Math. \textbf{187}, 1, 2018

\bibitem[RS18b]{RoSp18-b}
\textsc{Rodnianski}, I., \textsc{Speck}, J.:
\newblock \emph{On the nature of Hawking's incompleteness for the Einstein-vacuum equations: The regime of moderately spatially anisotropic initial data}
\newblock arXiv:1804.06825, 2018


\bibitem[Sp]{Sp12}
\textsc{Speck}, J.~:
\newblock \emph{The global stability of the Minkowski spacetime solution to the Einstein-nonlinear system in wave coordinates},
\newblock Anal. PDE, 7, 4, 2014, 771-901

\bibitem[Wi]{Wi82}
\textsc{Witten}, E.~:
\newblock \emph{Instability of the Kaluza-Klein vacuum},
\newblock Nuclear Physics B195, 1982, 481--492

\bibitem[Wy]{Wy17}
\textsc{Wyatt}, Z.~:
\newblock \emph{The Weak Null Condition and Kaluza-Klein Spacetimes},
\newblock  arXiv:1706.00026, 2017





 

\end{thebibliography}
\end{document}